\documentclass[useAMS,usenatbib]{mn2e}

\input{epsf}
\usepackage{graphicx}
\usepackage{rotating}

\usepackage{lscape}
\usepackage{multirow} 

\newcommand{\mic}{\,$\mu$m}
\newcommand{\kms}{\,km s$^{-1}$}
\newcommand{\dg}{$^{\circ}$}

\newcommand{\Msolar}{\,M$_{\odot}$}

\newcommand{\htwo}{H$_2$ 2.122\,$\mu$m}

\title[H$_2$ survey of Taurus-Auriga-Perseus]
{A shallow though extensive \htwo\ imaging survey of Taurus-Auriga-Perseus: 
I. NGC1333, L1455, L1448 and B1}
\author[C.J.~Davis et al.]
       {C.J.~Davis$^{1}$, 
        P. Scholz$^{1,2}$,
        P.~Lucas$^{3}$, M.D. Smith$^{4}$ and
        A. Adamson$^{1}$ \\
$^1$Joint Astronomy Centre, 660 North A'oh\={o}k\={u} Place,
        University Park, Hilo, Hawaii 96720, USA \\
$^2$Department of Physics and Astronomy, University of Victoria, 
        Victoria BC, Canada \\
$^3$Centre for Astrophysics Research, Science \& Technology Research Institute, 
        University of Hertfordshire, Hatfield AL10 9AB, UK \\
$^4$Centre for Astrophysics \& Planetary Science,
    School of Physical Sciences, The University of Kent,
    Canterbury CT2 7NR, England
}

\begin{document}

\date{Accepted 2007 ... ; 
      Received 2007 ... ; 
      in original form 2007 ... }

\pagerange{\pageref{firstpage}--\pageref{lastpage}} \pubyear{2008}

\maketitle

\label{firstpage}

\begin{abstract} 

We discuss wide-field near-IR imaging of the NGC1333, L1448, L1455 and
B1 star forming regions in Perseus.  The observations have been
extracted from a much larger narrow-band imaging survey of the
Taurus-Auriga-Perseus complex. These \htwo\ observations are
complemented by broad-band K imaging, mid-IR imaging and photometry
from the Spitzer Space Telescope, and published submillimetre CO J=3-2
maps of high-velocity molecular outflows. We detect and label 85 H$_2$
features and associate these with 26 molecular outflows. Three are
parsec-scale flows, with a mean flow lobe length exceeding 11.5\arcmin
. 37 (44\%) of the detected H$_2$ features are associated with a known
Herbig-Haro object, while 72 (46\%) of catalogued HH objects are
detected in H$_2$ emission.  Embedded Spitzer sources are identified
for all but two of the 26 molecular outflows.  These candidate outflow
sources all have high near-to-mid-IR spectral indices (mean value of
$\alpha \sim 1.4$) as well as red IRAC 3.6\mic -4.5\mic\ and IRAC/MIPS
4.5\mic - 24.0\mic\ colours: 80\% have [3.6]-[4.5]$>$1.0 and
[4.5]-[24]$>$1.5.  These criteria -- high $\alpha$ and red [4.5]-[24]
and [3.6]-[4.5] colours -- are powerful discriminants when searching
for molecular outflow sources.  
However, we find no correlation between $\alpha$ and flow length
or opening angle, and the outflows appear randomly orientated in each
region. The more massive clouds are associated with a greater number
of outflows, which suggests that the star formation efficiency is
roughly the same in each region.

\end{abstract}




\section{Introduction}

\begin{figure}
\epsfxsize=8.4cm
\epsfbox{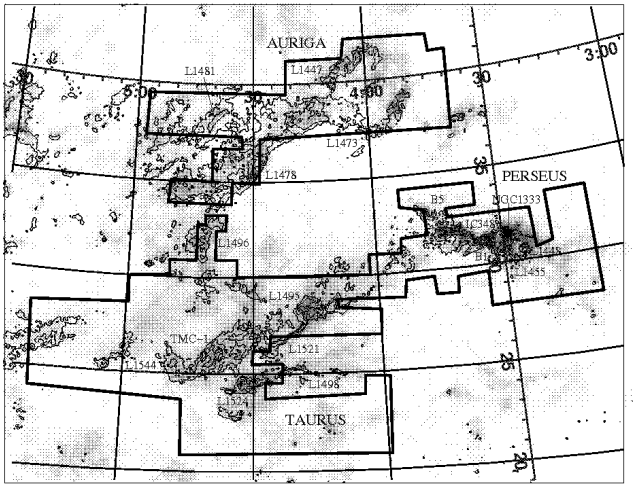}
\vspace*{0.3cm}
\caption[] {An overview of the Taurus-Auriga-Perseus complex. The
grey-scale shows integrated CO J=1-0 emission from a survey of the Milky Way
obtained with 1.2~m telescopes in Cambridge, MA and on Cerro Tololo,
Chile \citep{dam01}. The contours represent visual extinction, $A_{\rm v}$, 
constructed from optical star-counts using Digitised Sky Survey
data \citep{dob05}.  The black lines outline the complete region to 
be observed in \htwo\ emission and broad-band K. }
\label{over1}
\end{figure}

The U.K. Infrared Deep Sky Survey (UKIDSS) is a legacy survey being
conducted at the United Kingdom Infrared Telescope \citep{law07}.
UKIDSS comprises five surveys, ranging from very deep, narrow studies
of extra-galactic fields, to shallow, broad surveys at both low and
high galactic latitudes.  One such survey, the Galactic Plane Survey
(GPS), will by 2012 observe 1868 deg$^2$ of the northern and equatorial
galactic plane between -5\dg\ $<b<$ +5\dg , in J, H and K down to
limiting magnitudes of 20.0, 19.1 and 19.0, respectively
\citep{luc08}.  The GPS also includes a component covering
the Taurus-Auriga-Perseus (T-A-P) complex of star-forming clouds,
through narrow-band H$_2$ 1-0S(1) and broad-band K filters.  To date,
140 deg$^2$ of the T-A-P have been observed as part of the GPS.
Observations of an additional 40 deg$^2$ are currently being sought to
complete the proposed survey area, which is shown in Fig.~\ref{over1}.
The existing data are available from the WFCAM science archive
\citep[see][for details of how to access these data]{luc08} and, as
mosaicked images, from a dedicated database hosted by the Joint
Astronomy Centre\footnote{http://www.jach.hawaii.edu/UKIRT/TAP} using
an easy-to-navigate clickable map.  The use of the latter is
recommended for those wishing to quickly access imaging data to search
for H$_2$ emission features.

The goal of this paper is to present demonstration science from the
T-A-P H$_2$ survey.  We therefore focus on just the western end of
Perseus, comparing the narrow-band imaging data with photometry from
the Spitzer Space Telescope legacy programme {\em From Molecular Cores
to Planet-Forming Disks}, also known as {\em c2d} \citep{eva03}.  An
area covering almost 4 deg$^2$ (representing five WFCAM tiles)
has been searched for \htwo\ line
emission features.  The mid-IR data are then used to identify
candidate protostars for the molecular outflows found in H$_2$. We
also search for H$_2$ counterparts of the known Herbig-Haro (HH)
objects catalogued by \citet{wal05a}, and compare our images with the
CO outflow maps of \citet{kne00} and \citet{hat07}.  Our aim is to
establish the relative effectiveness of optical, near-IR and CO
surveys of outflows.
 
The Perseus cloud complex spans more than 30~pc at a distance of
between 200-350~pc \citep{dez99,cer90}.  The region was first mapped
in CO by \citet{sar79}; molecular line and extinction maps have since
been presented by \citet{rid06}.  The early search for Young Stellar
Objects (YSOs) and protostars from IRAS data by \citet{lad93} has
recently been superseded by the Spitzer surveys of
\citet{jor06}, \citet{lad06} and \citet{gut08}.  Perseus harbours two 
particularly active star forming regions, NGC1333 -- centred on the
bright and very well studied HH complex HH~7-11
\citep{hod95,bal96,chr00}-- and IC~348, a rich, embedded stellar
cluster \citep{luh98,luh05,lad06}.  Here we focus only on NGC1333 and
surrounding clouds -- a region we refer to as ``Perseus-West'' --
although the full extent of Perseus will be imaged in H$_2$ as part of
the T-A-P survey (Fig.~\ref{over1}), and in JHK by the UKIDSS Galactic
Clusters Survey
\citep{law07}.


\begin{figure}
\epsfxsize=8.6cm
\epsfbox{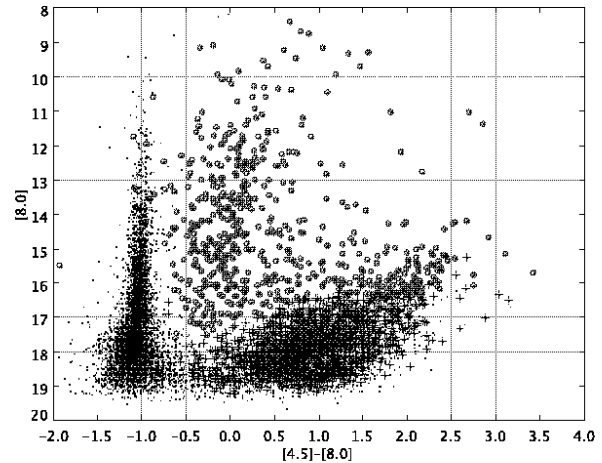}
\caption[] {Colour-magnitude diagram of c2d Perseus data. Black dots
represent the entire c2d Perseus catalogue; red circles mark those
identified as YSOs, while blue crosses mark those labelled ``red''
in the catalogue.}
\label{c2dcmag}
\end{figure}

\section{Observations}

\subsubsection{WFCAM data}

The near-IR wide-field camera WFCAM \citep{cas07} contains four
Rockwell Hawaii-II (HgCdTe 2048x2048) arrays spaced by 94\% in the
focal plane.  The pixel scale measures 0.40\arcsec . To observe a
contiguous square field on the sky covering 0.75~deg$^2$, a region
known as a WFCAM ``tile'', observations at four positions are
required. At each of these four positions eight exposures are obtained
in each filter, with a 20~sec exposure time in H$_2$ and a 5~sec
exposure time in K.  The images are shifted by an integral multiple of
0.2\arcsec\ so that in the final, mosaicked images the pixel scale is
0.2\arcsec ; in this way the sub-arcsecond seeing typically obtained
at Mauna Kea is fully sampled and bad pixels may be compensated for.

All WFCAM data are initially reduced by the Cambridge Astronomical
Survey Unit (CASU) and are distributed through a dedicated archive
hosted by the Wide Field Astronomy Unit (WFAU). The CASU reduction
steps are described in detail by \citet{dye06} and \citet{irw08};
astrometric and photometric calibrations are achieved using 2MASS
\citep{dye06,hew06}.  Data are then downloaded in bulk from WFAU and 
combined into tiles using astrometric information stored in the file
headers.  As part of this process, residual sky/background structure
is removed by fitting a coarse surface to each image
\citep[as described by][]{dav07}.  Currently data for 140~deg$^2$ are 
available; H$_2$ and K-band mosaic tiles are available to all
interested parties from a database at the Joint Astronomy Centre
(http://jach.hawaii.edu/UKIRT/TAP) as Rice-compressed fits files.

\subsubsection{Spitzer data}

The Spitzer Space Telescope observations discussed in this paper were
obtained with the IRAC and MIPS cameras, which are described by
\citet{faz04} and \citet{rie04}, respectively.  
IRAC data in bands 1,2,3 and 4 (at 3.6, 4.5, 5.8 and 8.0\mic) and
MIPS band 1 data (at 24\mic ) were obtained in September 2004 as part
of the Spitzer legacy programme {\em c2d}
\citep{eva03}.  The data have been presented
by \citet{jor06}; guaranteed time observations of NGC1333
are presented by \citet{gut08}. Additional data on L1448 were 
obtained on February 25, 2005 by \citet{tob07}.

The mid-IR photometry used here were obtained directly from the {\em
c2d} archive (http://peggysue.as.utexas.edu/SIRTF/), which includes
data extracted from the third delivery of IRAC and MIPS data.  The
procedures used during source extraction are described by
\citet{eva05} and \citet{har07}.  For Perseus, the full catalogue of
sources detected in at least one IRAC band contains over 120,000
objects. IRAC and MIPS band 1 (24\mic ) data have been combined with
2MASS JHK photometry by the {\em c2d} group to distinguish candidate
Young Stellar Objects (YSOs) and protostars from field stars and
background galaxies \citep[see also][]{jor06}. A spectral index,
$\alpha = d log (\lambda F(\lambda))/d log \lambda$, calculated
between 2.2\mic\ and 24\mic , is included in the catalogue for all
source with at least two photometry points.  In this paper we
over-plot the positions of YSOs and protostars identified in this way,
as well as the ``red'', ``red1'' and ``red2'' sources (targets with a
high MIPS band 1 to IRAC flux ratio, or those with rising spectral
energy distributions [SEDs] detected in only MIPS1 or MIPS1 and IRAC
band 4) and ``PAH-em'' and ``star+dust'' sources (targets with IRAC
colours expected for PAH emission features; these may include
background galaxies but also Class II sources). These latter sources
are arguably also candidate protostars and/or outflow sources.  

Note that both this paper and the c2d catalogue refer to all young
stars as YSOs, regardless of whether they are Class 0/I protostars or
Class II/III young stellar objects.

Sample SEDs, plotted between 1.2\mic\ and 24\mic , are presented for
all source types by \citet{eva05}.  In Fig.~\ref{c2dcmag} we plot a
Spitzer colour-magnitude diagram for sources in Perseus West from the
c2d catalogue. There are 66,535 sources in the catalogue, of which 4,352
are ``red'' sources and 610 YSOs. Clearly, most of the ``red'' sources, 
though red in the two IRAC bands, have a low 8.0\mic\ flux.  In the most
deeply embedded objects this may be caused by silicate absorption or
the intrinsic shape of the SED (Class 0 sources have been shown to
possess a [3.6]-[4.5]/[5.8]-[8.0] ratio that is $>$1; J\o rgensen et
al. 2006).  Even so, the ``red'' sources may also include a modest
population of background galaxies \citep{har06,jor06}.

\subsubsection{JCMT data}

\begin{figure*}
  \epsfxsize=15.0cm
  \epsfbox{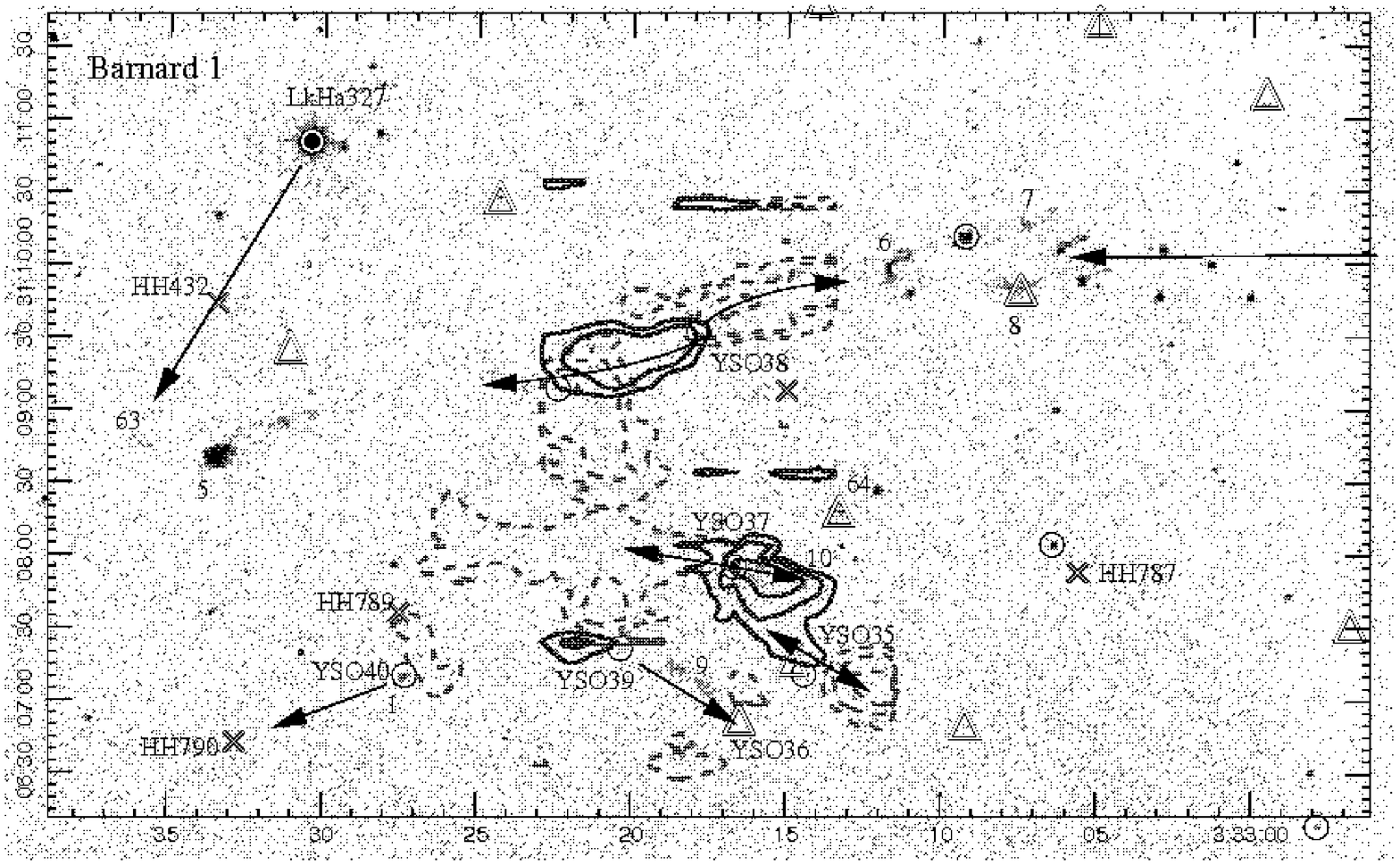}
\caption[]
{Logarithmically-scaled \htwo\ image of B1 with the locations of HH objects marked with
crosses, Spitzer-identified YSOs marked with circles and
Spitzer-identified red sources marked with triangles. J\o rgensen
embedded YSOs are labelled. Contours of high-velocity CO J=3-2 emission
\citep[from][]{hat07} are over-plotted: integrated antenna temperature
is plotted from 9 to 15\kms\ (dashed lines) and -5 to 4\kms\
(full lines). Contour levels are 1,2,4,8 K\kms. 
H$_2$ features are numbered for reference (see Table~\ref{jets}). 
Outflows are indicated with arrows.}
\label{h2B1}
\end{figure*}

\begin{figure*}
  \epsfxsize=17.5cm
  \epsfbox{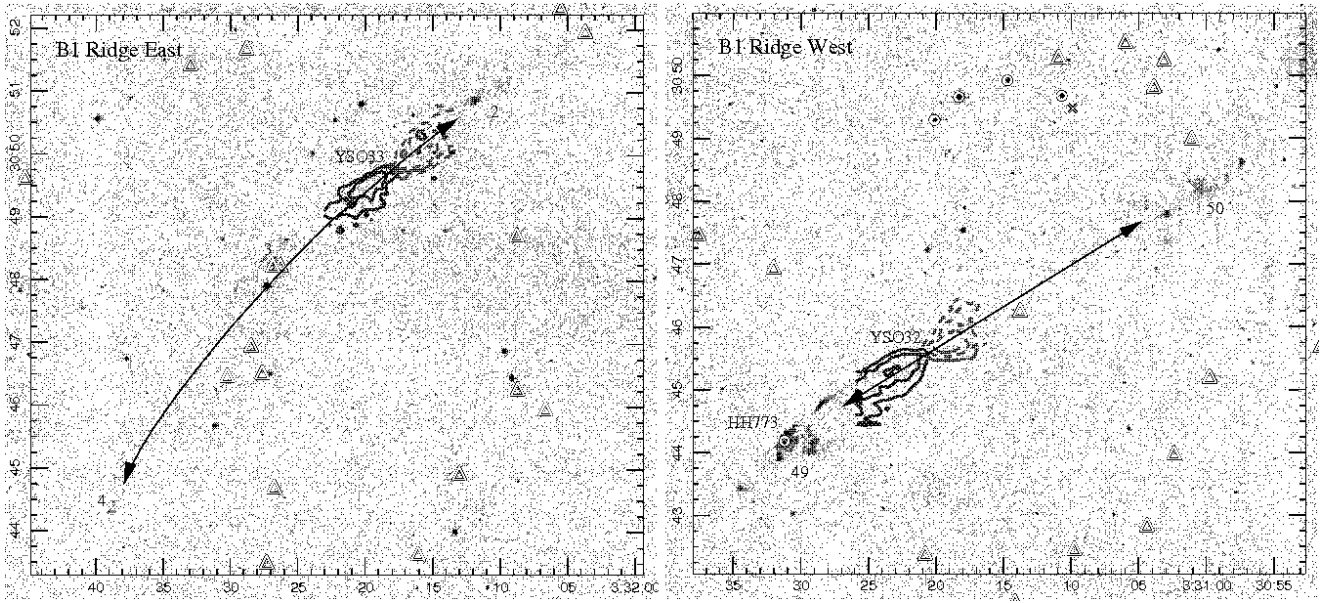}
\caption[]
{\htwo\ images of two adjacent regions in the B1 ridge. Contours of
high-velocity CO 3-2 emission \citep [from][]{hat07} are over-plotted
from 9 to 15\kms\ (full lines) and from -5 to 4\kms\ (dashed
lines). Contour levels are 0.5,1,2 K\kms\ (left panel) and 1,1.5,2
K\kms\ (right panel). See Fig.\ref{h2B1} for further details.}
\label{h2B1r}
\end{figure*}

\begin{figure*}
  \epsfxsize=17.5cm
  \epsfbox{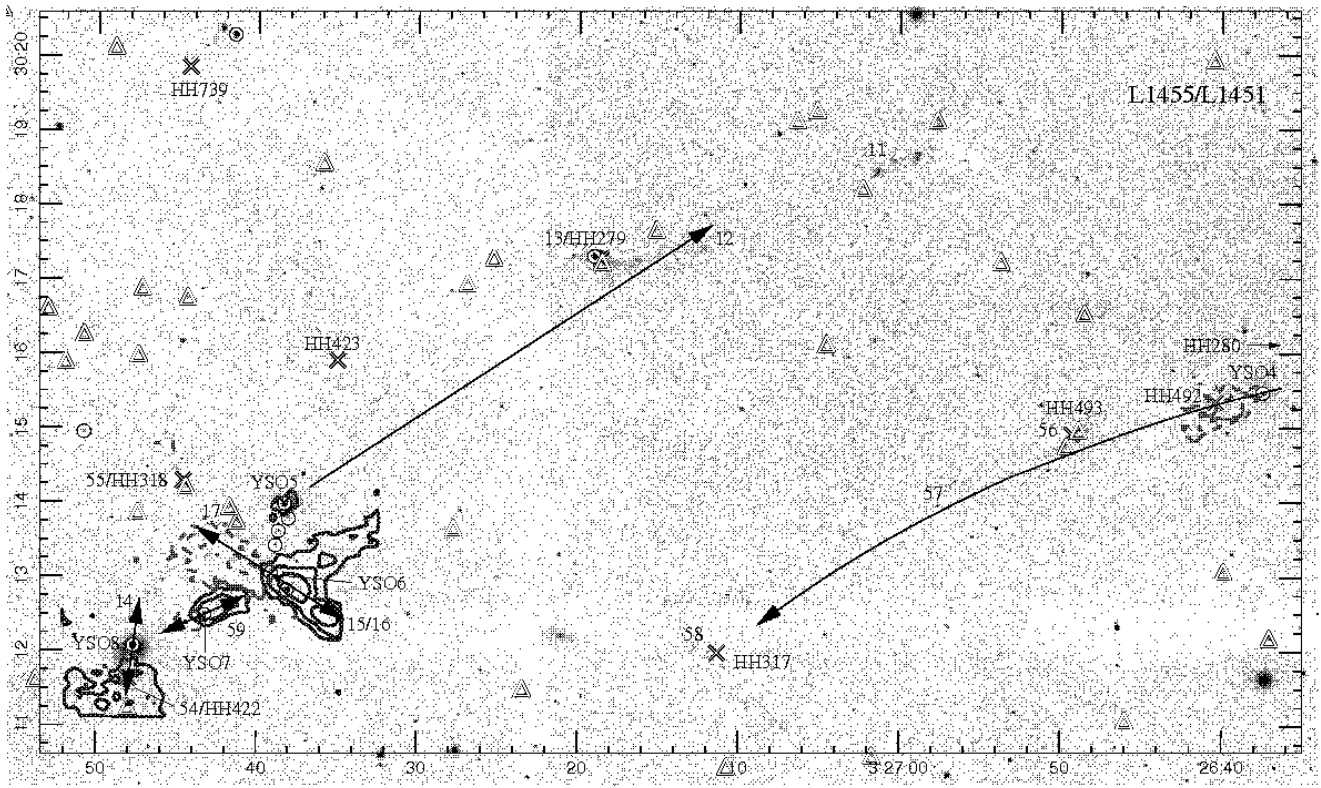}
\caption[]
{\htwo\ image of L1455 with L1451 to the west.
Contours of high-velocity CO 3-2 emission
\citep [from][]{hat07} are over-plotted from 6 to 15\kms\ (full lines)
and from -5 to 3\kms\ (dashed lines).
Contour levels are 0.5,1,2 K\kms.
See Fig.\ref{h2B1} for further details.}
\label{h2L1455}
\end{figure*}

\begin{figure*}
  \epsfxsize=17.5cm
  \epsfbox{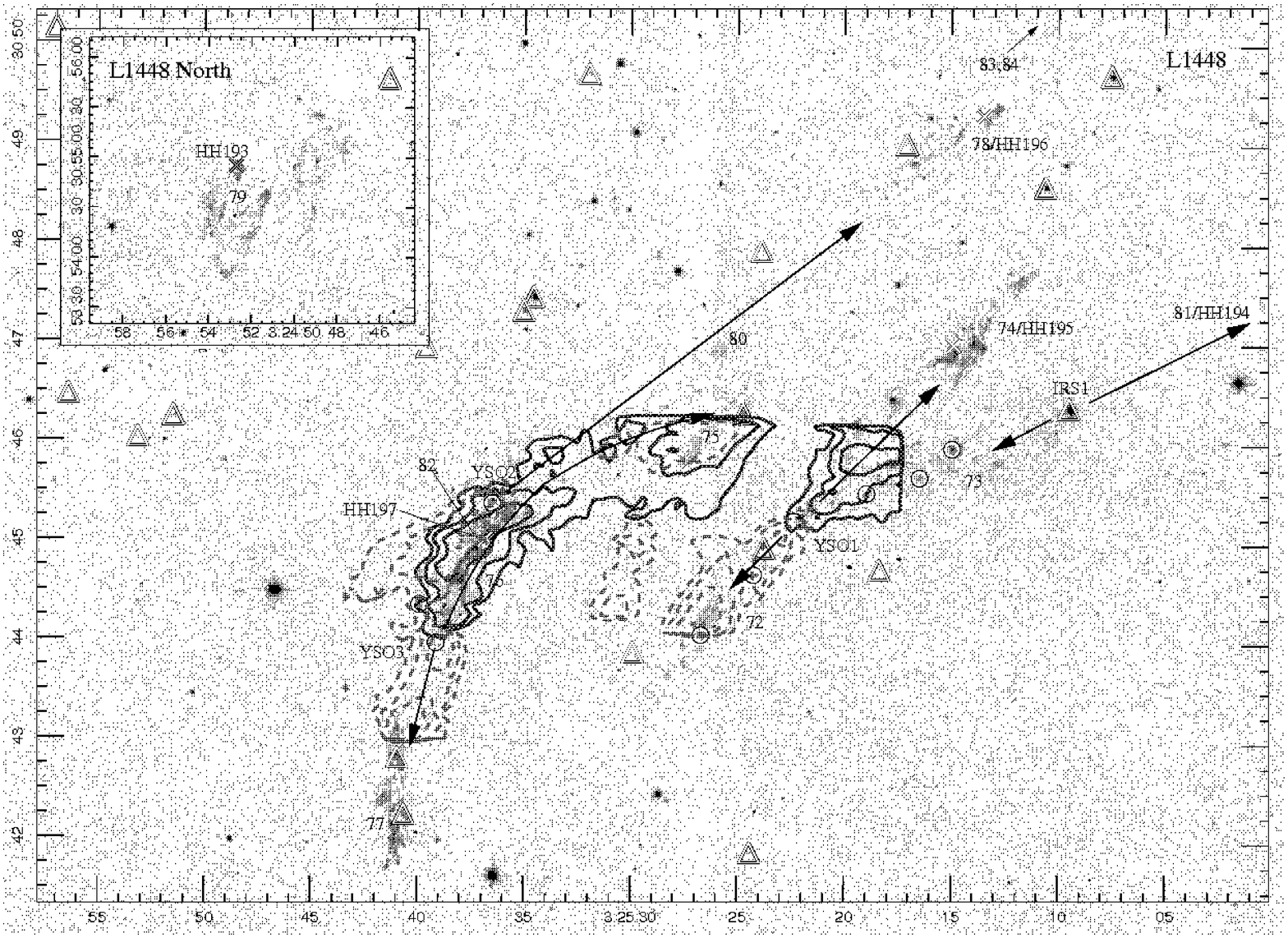}
\caption[]
{\htwo\ image of L1448. Contours of high-velocity CO 3-2 emission
\citep [from][]{hat07} are over-plotted from 8 to 15\kms\ (full lines)
and from -5 to 0\kms\ (dashed lines).
Contour levels are 1,2,4,8 K\kms.
See Fig.\ref{h2B1} for further details.}
\label{h2L1448}
\end{figure*}

\begin{figure*}
  \epsfxsize=17.0cm
  \epsfbox{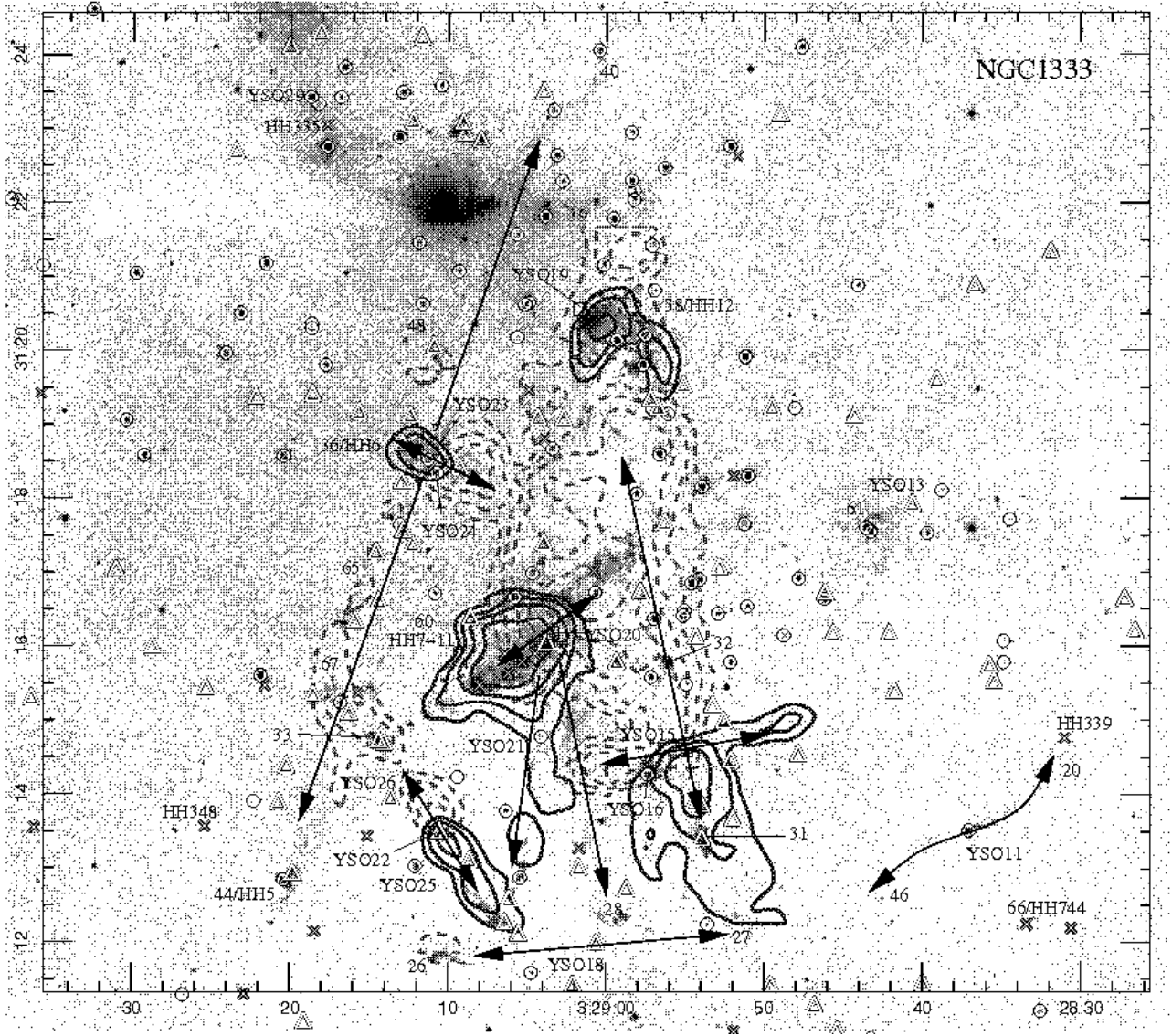}
\caption[] 
{\htwo\ image of NGC1333. Open squares mark SCUBA cores
without c2d-identified YSOs; the other symbols are as in
Fig.\ref{h2B1}. High-velocity CO 3-2 contours
\citep [from][]{kne00} are over-plotted, from 12 to 18\kms\ (full lines)
and from -5 to 3\kms\ (dashed lines). Contour levels are 1.5,3,6,12 K\kms . }
\label{h2NGC1333}
\end{figure*}

The H$_2$ emission in outflows is known to be produced in molecular
shocks formed at the interface between a jet and the ambient medium.
Thus, the H$_2$ knots and bow shocks imaged with WFCAM are expected to
be closely related to the high-velocity molecular flow lobes often
mapped at (sub)millimetre wavelengths in rotational CO emission.  In
Figs.\ref{h2B1} to \ref{h2NGC1333} we therefore present CO J=3-2
observations of high-velocity molecular outflows in Perseus-West taken
from the literature.  These data were acquired at the James Clerk
Maxwell Telescope (JCMT) and have been published and discussed in
detail by \citet{kne00} and \citet{hat07}.  The JCMT beam width
measures 14\arcsec\ at the 345.796~GHz frequency of the
J=3$\rightarrow$2 transition.  Contour plots are presented in units of
antenna temperature.

The maps cover only very limited regions in each star-forming cloud,
though many of the flows discussed below have been observed.  More
extensive maps are being planned as part of the {\em Gould Belt}
legacy survey at the JCMT \citep{war07}.


\begin{figure*}
  \epsfxsize=17.0cm
  \epsfbox{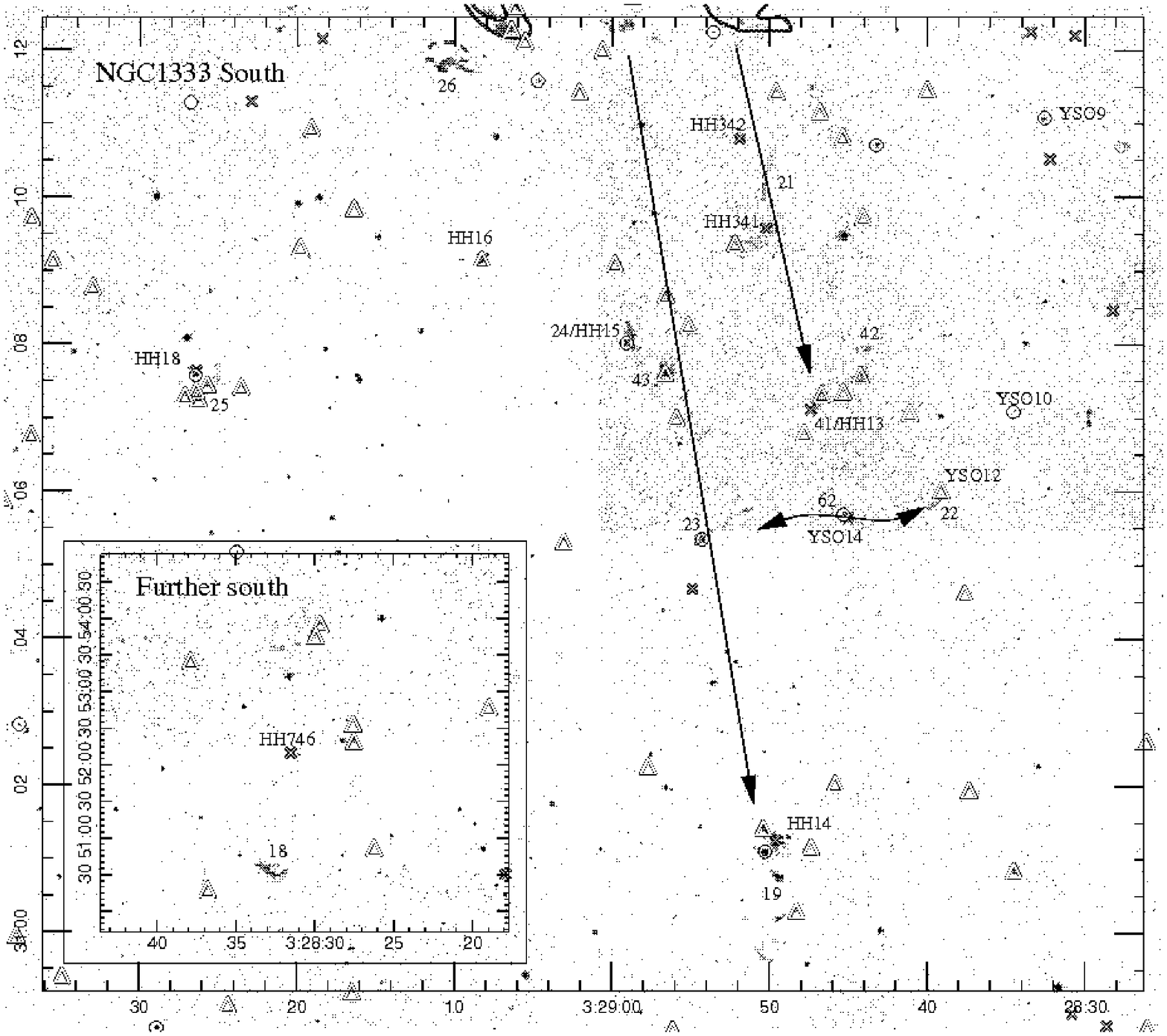}
\caption[] 
{\htwo\ image of the region to the south of
NGC1333. Labelling and symbols are as in Fig.\ref{h2B1}.}
\label{h2NGC1333S}
\end{figure*}


\section{Results \& Discussion}

\subsection{H$_2$ imaging of Perseus-West}

The five WFCAM tiles used in this paper cover the western portion of
Perseus, which includes NGC1333, L1455, L1448, and Barnard 1
(B1). WFCAM H$_2$ images of regions where outflows were detected are
shown in Figs~\ref{h2B1}--\ref{h2NGC1333S}. The positions of
protostars and YSOs identified in the Spitzer c2d survey are
overplotted in each figure.  Circles indicate sources that were given
the ``YSO'' designation in the c2d catalogue, while triangles mark
sources classified as ``red'', ``red2'', ``PAH-em'', or
``star+dust''. \citet{jor07} have compared the c2d data with SCUBA
observations at 850\mic . They present a catalogue of 49 ``embedded''
YSOs; all are MIPS detections that are either red in [3.6]-[4.5] and
[8.0]-[24] colours or are located within 15\arcsec\ of SCUBA peaks.
They also include a few concentrated SCUBA cores with no MIPS
counterparts. We label these sources as ``YSO'' with the numbers from
J\o rgensen et al.'s catalogue (their Table 3) and list those sources
located in the Perseus-West region in Table~\ref{outflows}.

The locations of Herbig-Haro (HH) objects found by \citet{wal05a} in
their wide-field optical imaging, as well as HH objects that were
previously known, are also marked on our figures. Note
that these positions are often rather vague, representing the location
of what is usually an extended object or diffuse feature.

In the online Appendix we present UKIRT-WFCAM/Spitzer-IRAC
colour-composite images of each region, and discuss the outflows
marked in Figs.~\ref{h2B1}--\ref{h2NGC1333S} in some detail, noting in
particular the association of H$_2$ emission-line features with HH
objects, molecular CO outflows, dense cloud cores and the embedded
source identified through Spitzer photometry.  H$_2$ knot parameters
(flow position angles, flow lengths, associated HH objects, etc.) are
listed in Table~\ref{jets}, where we also give the embedded YSO that
is likely to be driving each flow. Many of these knots or groups of
knots are known HH objects or parts of outflows imaged in less
extensive surveys. We therefore also list the names of associated HH
objects and previous H$_2$ outflow names in Table
\ref{jets}.

\subsection{On the merits of combining optical, infrared, and
submillimetre observations of outflows}

Overall, the combined WFCAM H$_2$ imaging, Spitzer imaging and
photometry, and CO outflow observations provide a clear and rather
complete picture of dynamic activity in Perseus-West.   Our
WFCAM observations demonstrate that even shallow H$_2$ imaging is an
effective tool for finding outflows from the youngest YSOs.

Of the 26 molecular outflows in our WFCAM tiles, 24 of them (92\%) are
driven by sources identified as embedded YSOs by \citet{jor07}; only
9\% of the 85 H$_2$ features have no obvious CO
flow counterpart or J\o rgensen YSOs.  The two outflow sources not
identified by J\o rgensen et al. as embedded YSO, LkH$\alpha$327 and
L1448-IRS1, were probably missed because of saturation in the Spitzer
bands.  Neither source coincides with a SCUBA core \citep{hat05}, so
these bright sources have probably already cleared their cores.  Of
the 38 J\o rgensen YSOs in our field, 24 (63\%) drive
outflows. Clearly, the criteria used by \citet{jor07} (MIPS/IRAC
colours [3.6]-[4.5]$>$1 and [8.0]-[24]$>$4.5 and/or associated SCUBA
cores) is not only very good at identifying YSOs, but also at
identifying YSOs that drive molecular outflows. 

Shock-excited outflow features can of course be detected at both
optical and infrared wavelengths, by virtue of their forbidden
emission lines ([OI], [SII], [OIII], etc.), hydrogen recombination
lines (H$\alpha$), or -- in the near-IR and mid-IR -- molecular
hydrogen ro-vibrational lines. However, the longer-wavelength
transitions will be more effective in regions of high
extinction. Also, these lines derive from very different gas
components: the optical lines typically trace hot, dense,
partially-ionised, high-velocity gas (T$\sim$10,000\,K; flow
velocities approaching 100-200~km s$^{-1}$) while the H$_2$ lines
trace low-excitation, shocked molecular gas
at much lower velocities (T$\sim$500-2,000\,K; v$\sim$10-50~km
s$^{-1}$). It is therefore perhaps not surprising that, of the 158 HH
objects from \citet{wal05a} in our field, only 72 (46\%) of them were
detectable in \htwo\ emission. Similarly, only 37 (44\%) of the 85
labelled H$_2$ features are associated with an HH object. Many of the
HH objects are faint in the optical, so in a few cases the
lack of a detection in near-IR H$_2$ emission may be due to the modest
sensitivity of the WFCAM survey.  However, many of the HH objects are
located around the periphery of each star-forming cloud (see for
example the comparison of HH and H$_2$ positions in
Fig.~\ref{h2NGC1333}), where the molecular gas density may be too low
to facilitate H$_2$ excitation.  Clearly, a colour-composite image
comprised of optical ([SII] or H$\alpha$), near-IR 2.12\mic\ and
mid-IR 4.5\mic\ imaging would be ideal for illustrating the
high-excitation atomic/ionised (HH), hot molecular (ro-vibrational
H$_2$), and warm molecular (pure-rotational H$_2$) flow components
simultaneously, as well as the colours of white/blue foreground stars
through to the reddest protostars.  Such a comparison is obviously now
possible with ground-based optical imaging, WFCAM survey data, and
Spitzer observations, for a number of well-known star forming regions.


\begin{figure*}
  \epsfxsize=16.5cm
  \epsfbox{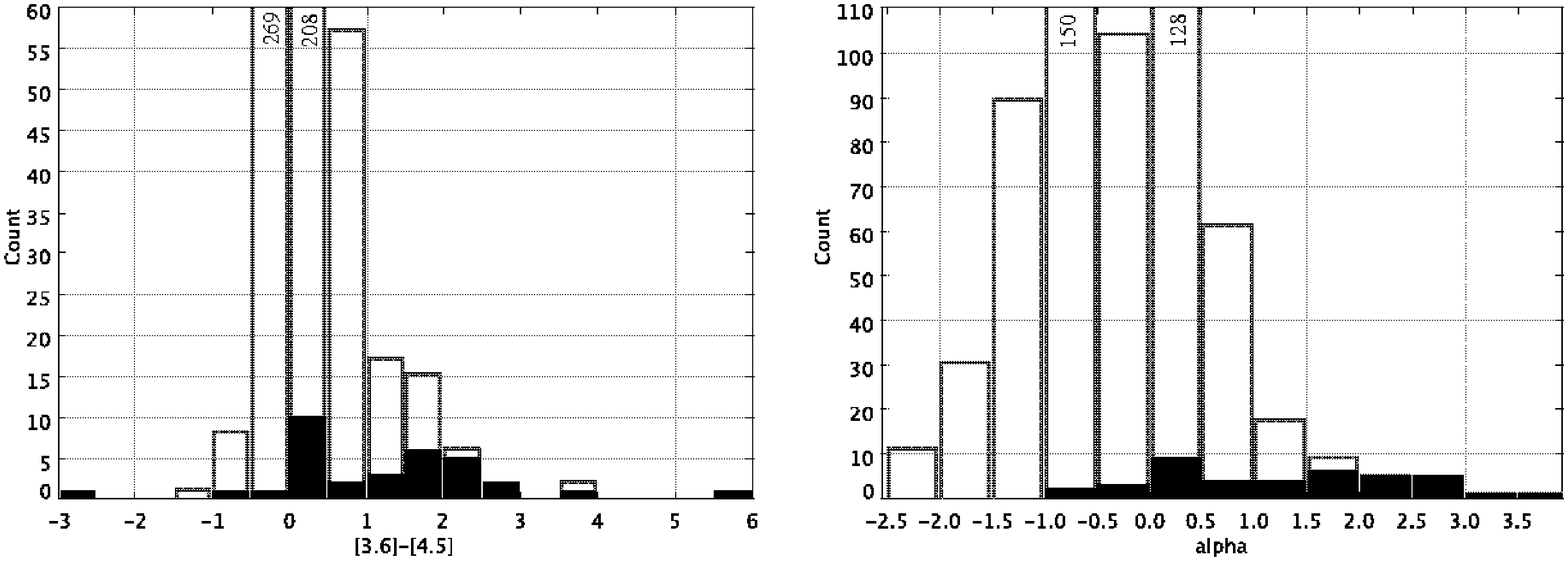}
\caption[] {Histograms of IRAC [3.6]-[4.5] colour and source spectral
index, $\alpha$, for YSOs in Perseus. Open bars represent all c2d
identified YSOs; filled bars are sources with identified outflows.}
\label{histograms}
\end{figure*}

\begin{figure*}
  \epsfxsize=16.5cm
  \epsfbox{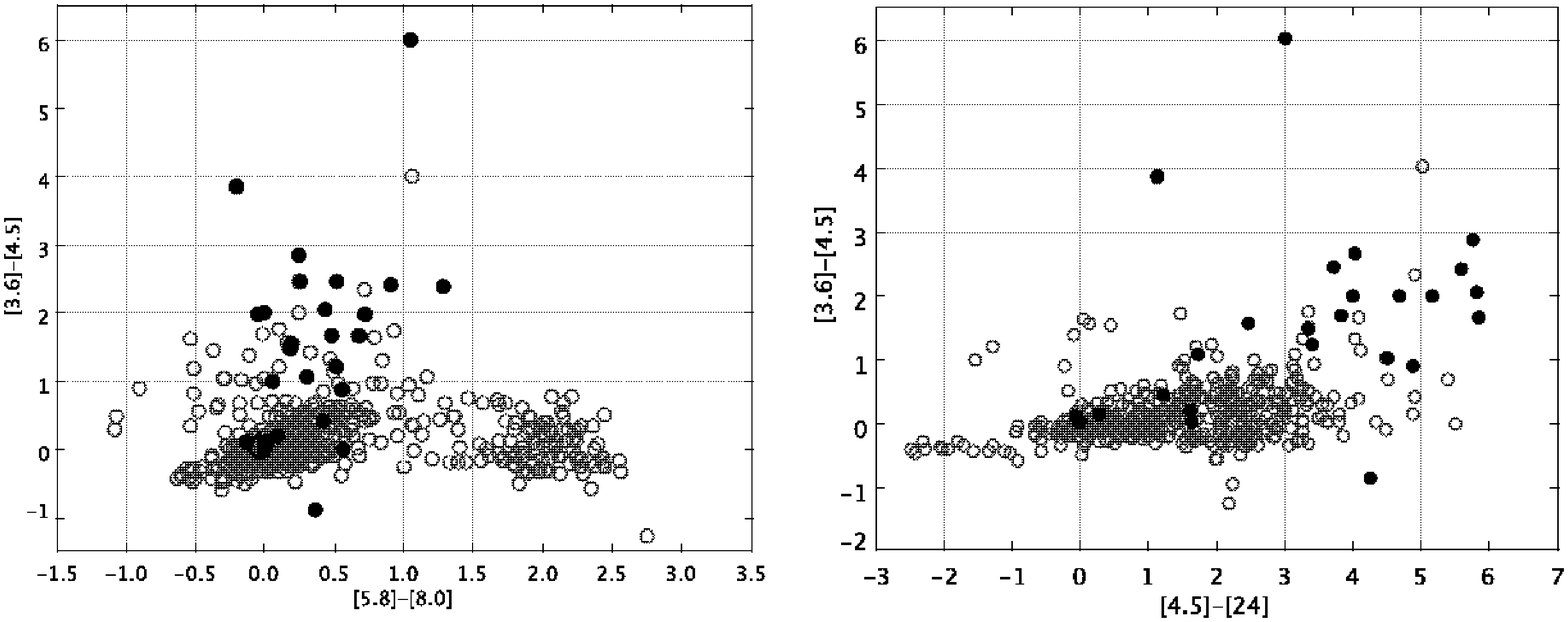}
\caption[] {Spitzer colour-colour diagrams
for c2d YSOs in Perseus (open circles). Filled circles 
represent sources with identified outflows.}
\label{ccandcmag}
\end{figure*}


\subsection{General H$_2$ flow characteristics}

Of the 26 H$_2$ outflows identified in column 2 in Table~\ref{outflows}, 20
(77\%) exceed an arcminute in mean length; three are parsec-scale
flows, with a mean lobe length exceeding 11.5\arcmin\ (the outflows
driven by YSOs 2, 5, and 20c).  Longer flows might be expected from
more evolved sources, and indeed YSO~20c and YSO~5 do have flat spectral
indices.  However, YSO~2 has a steeply rising SED ($\alpha \sim 2.3$).
This high value of $\alpha$ may in part be due to excess emission from
lines in some of the Spitzer bands (discussed below).  However, when
we plot flow length against Spitzer spectral index we find no
correlation between these two parameters.  This suggests that even the
most deeply embedded H$_2$ jet source are old enough to produce
parsec-scale flows.  It seems likely that jet length, when derived
from H$_2$ observations, is more sensitive to the environment
(specifically the presence of ambient molecular gas that may be
shock-excited into emission) than the ``age'' of the jet central
engine.

Can we say anything about flow opening angles from our sample of jets
in Perseus-West?  High-resolution CO maps of outflows seem to indicate
that Class 0 sources produce more collimated flows than their more
evolved Class I counterparts \citep[e.g.][]{lee02,arc06}.  H$_2$
emission features (bow shocks) often envelope the outer edges of CO
flow lobes, so one might expect to see a similar trend in H$_2$ data.
However, shock-excited H$_2$ emission has a very short cooling time
(of the order of a few years), so H$_2$ features are often short-lived
and confined to a small percentage of the flow surface area.
Moreover, close to the central source H$_2$ emission features tend to
be associated with the central, on-axis jet rather than the swept-out
cavity walls. Defining flow opening angles from H$_2$ images is
therefore rather difficult. Even so, in Table~\ref{outflows} we list
the mean opening angle, $\theta$, for well-defined flows with
multiple, bright H$_2$ emission features.  $\theta$ is measured from
cones in each flow lobe that are centred on the outflow source and
which include all H$_2$ line-emission features.  A plot of $\theta$
against source spectral index $\alpha$ (not shown) reveals no
correlation between these two parameters.  At best, we can only say
that most of the flows in Perseus-West appear to be well collimated,
as is expected for jets from young, embedded (high-$\alpha$)
sources. The opening angles listed have a mean of 12.5\dg\ and a
standard deviation of only 4.1\dg. The associated sources have a
relatively broad range in $\alpha$: -0.87 to 3.79.

Many of the H$_2$ jets in Perseus appear to be curved or possibly even
precessing. If both lobes curve in the same direction (H$_2$ features
2--4 in the B1 ridge and 76/77 in L1448 are good examples) the
flow curvature may be due to the motion of the source through the
ambient medium.  In such cases the source would need a proper motion
that is a sizable fraction (at least a few percent) of the jet
velocity. Source tangential velocities of the order of 1-10\kms\ would
be required. Alternatively, the {\em apparent} curvature may simply be
due to excitation of H$_2$ along only one side of an oval cavity,
perhaps because of enhanced ambient gas densities or an ambient
density gradient across the width of the flow lobe.

Precession or jet ``meandering'' has been identified in a number of flows
\citep[e.g.][]{lop95,hea96,dav97,hod05}.  Precession may result from the
misalignment of binary orbital and circum-binary disk axes
\citep{ter99,bat00}, although the disk should be forced into alignment
with the orbital plane in roughly one precession period, which should
be about 20-times longer than the binary orbital period \citep{bat00}.
Given this time limitation, jet precession is expected from only the
most embedded and therefore potentially very youngest sources.
Notably, YSOs 11, 14 and 38, the three precessing jet
candidates in Perseus-West, all have high values of $\alpha$.

The H$_2$ flows from YSOs 11, 14 and 38 appear to have precessed
through roughly half a turn.  The dynamical age of each flow should
thus be a fair estimate of the precession period, to within a factor
of $\sim$2 \citep{hod05}.  For a canonical jet velocity of 100~\kms\
\citep[consistent with jet proper motions, e.g.][]{eis94}, from the
flow lengths listed in Table~\ref{outflows}, flow ages of 7300~yrs,
1400~yrs and 2600~yrs are estimated. 
Hence, a very crude range for the precession period of
these remarkable flows is 2800-14,600~yrs.


\subsection{Photometric characteristics of outflow sources}

Fig.~\ref{histograms} demonstrates how the outflow sources differ from
the bulk of the c2d-catalogue YSOs in Perseus. The outflow sources,
plotted with black filled bars, are generally much redder and have a
steeper near-IR/mid-IR SED than the general YSO population. The flow
sources identified in this paper have a mean spectral index ($\alpha$)
of 1.39$\pm$1.14; the c2d YSOs, which are predominantly Class II
T~Tauri stars, have a mean $\alpha$ of -0.32$\pm$0.89. (In the c2d
catalogue, in regions with $A_{\rm v}<5$ where sources are
mostly background main sequence stars, the average value of
$\alpha$ is -2.8: Evans et al. 2005.)

The outflow sources also occupy a relatively distinct region in
colour-colour space (Fig.~\ref{ccandcmag}).  They possess a high
[4.5]-[24] IRAC/MIPS colour, as predicted for young sources with massive
accretion disks \citep{whi04,all04}. However, they also exhibit
positive [3.6]-[4.5] colours, unlike the bulk of the c2d YSOs, where
the near-zero [3.6]-[4.5] colours are due to the intrinsic shape of
the YSO SED. In Fig. \ref{ccandcmag}, 80\% of the outflow sources that
have measured 3.6\mic~ and 4.5\mic~ magnitudes have a [3.6]-[4.5]
colour greater than 1. This is compared to only 11\% of the total c2d
catalogue of YSOs in Perseus.  Of the YSOs with a [3.6]-[4.5] colour
greater than 1.0, 18\% have outflows; only 0.4\% of YSOs with
[3.6]-[4.5]$<$1.0 drive outflows.  All factors point to the
association of outflows traced in H$_2$ with the youngest Class 0/I
protostars.

\begin{figure}
  \epsfxsize=8.4cm
  \epsfbox{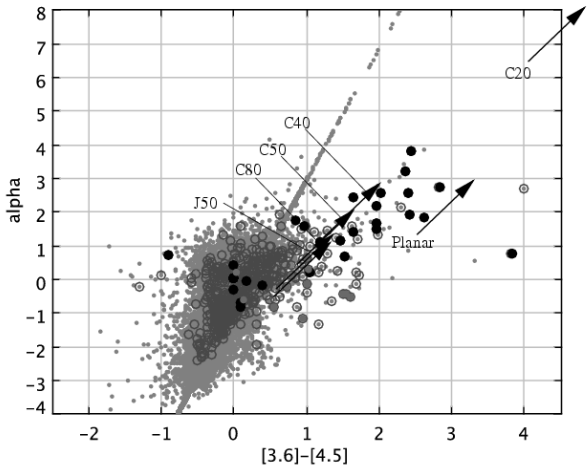}
\caption[] {Plot of spectral index ($\alpha$) against [3.6]-[4.5]
colour. Grey dots represent the entire c2d Perseus catalogue, red open
circles represent c2d identified YSOs, and black filled circles represent
outflow sources. Green circles mark c2d YSOs that may actually be
misidentified H$_2$ knots. The arrows show the predicted positions for
H$_2$ emission from a Planar shock, a J-type bow shock and four C-type bow
shocks \citep{smi05}. The arrows reflect an increase in
K-band extinction of 5~magnitudes.}  
\label{alphacolor} 
\end{figure}

In Fig.~\ref{alphacolor} we plot $\alpha$ against [3.6]-[4.5] colour
for the entire Perseus c2d catalogue.  Sources with values of $\alpha$
derived from just the two Spitzer bands plotted on the x-axis
(i.e. those detected in only bands 1 and 2) lie along a narrow,
diagonal strip.  Sources detected in multiple Spitzer and/or 2MASS
bands exhibit some scatter about this strip, since $\alpha$ is derived
from photometry points that are modified by various absorption
features (ice, silicate dust, etc.) and emission lines (H$_2$, [FeII],
[NeII], etc.), as well as the intrinsic shape of the YSO SED
(i.e. that of an embedded source with a disk and probably also a
disk-hole).  The Spitzer YSOs tend towards higher values of $\alpha$
and redder [3.6]-[4.5] colours, though they are largely
indistinguishable from the other sources (field stars, background
galaxies, etc.) in this plot. However, the outflow sources stand out
as having excess emission at 4.5\mic\ and thus red [3.6]-[4.5]
colours.  This is partly due to the inclusion of emission lines
associated with infall and outflow in their integrated 4.5\mic\
fluxes.  The 4.5\mic\ Spitzer band covers the H$_2$ pure-rotational
0-0S(9), S(10) and S(11) lines at 4.69, 4.41 and 4.18\mic , as well as
the CO v=1-0 band between 4.45-4.95\mic\ and the H{\sc I} Br$\alpha$ line at
4.052\mic , both of which may be excited in accretion flows.

Some of the sources identified in the c2d catalogue as YSOs, based on
their Spitzer colours and/or spectral index, will in fact be molecular
hydrogen emission-line knots. Shock models developed by \citet{smi05}
predict [3.6]-[4.5] colours in excess of $\sim$0.5, as well as
positive values for $\alpha$ (when derived from all four IRAC bands).
These predictions are supported by recent Spitzer observations of
HH~46/47, where the jet and bow shocks are detected in all four IRAC
channels \citep{vel07}
-- note that there are pure-rotation H$_2$ emission lines in all
of the IRAC bands. 

In Fig.~\ref{alphacolor} we plot the [3.6]-[4.5]
colours and corresponding spectral indices for six shock models.
The arrows indicate the effect of 5 magnitudes of K-band extinction
(reddening moves the points to the upper-right in the figure).  The
arrow labelled ``Planar'' represents a slab of molecular gas at a
temperature of 2000~K with a density of $10^5$~cm$^{-3}$.  Arrows
labelled C80, C50, C40 and C20 are C-type bow shocks: all have the same
shape and post-shock density ($10^5$~cm$^{-3}$), differing only in
shock velocity (80\kms , 50\kms , 40\kms\ and 20\kms\ respectively).
Decreasing shock velocity pushes the C-bow data points toward the
upper-right in Fig.~\ref{alphacolor}.  Higher values of $\alpha$ and
redder [3.6]-[4.5] colours are produced by a relative increase in
emission from the lower-energy pure-rotational H$_2$ lines (0-0S(4)
and S(5) in IRAC band 4 and 0-0S(6) and S(7) in band 3) over the
higher energy 0-0 lines (S(9)-S(11) in band 2 and S(13) and above in
band 1) and ro-vibrational lines (1-0O(5) and O(7) in band 1) at shorter
wavelengths.  The fastest C-type bow, C80, has a dissociative cap and
is much like the J-type bow shock.  It is therefore not too surprising
that it coincides almost precisely with the 50\kms\ J-type bow model
arrow, J50, in Fig.~\ref{alphacolor}.

The shock model data points overlap quite significantly with the
outflow sources in Fig.~\ref{alphacolor}.  The c2d ``YSOs'' that are
most likely to be H$_2$ knots, features 7, 12, 13, 16, 38, 49 and
72 (the green circles in Fig.~\ref{alphacolor}) have relatively low
values of $\alpha$ and therefore appear to be associated with either
fast C-type or J-type bow shocks.  The low spectral indices measured
for the observed knots could be caused by CO bandhead emission in IRAC
band 2 or the inclusion of 24\mic\ data when calculating $\alpha$ for
the observed knots.  We do not include this data point in the model
points, since there are no H$_2$ lines in the MIPS 24\mic\ bandpass.
However, weak forbidden emission from [FeII] at 24.51\mic\ and
25.98\mic\ \citep[like that seen in HH~46/47; ][]{vel07} could result
in the ``weak'' detection of an outflow feature in this band in the c2d
catalogue, and hence a lower value of $\alpha$ based on a fit to IRAC
and MIPS data.  CO emission is unlikely to contribute to observed
outflow photometry data since very high densities, in excess of
10$^{6}$~cm$^{-3}$, are required to produce appreciable flux from this
molecule.  However, under such circumstances (seen perhaps in high-mass star
forming regions) CO could account for as much as 10\% of the observed
emission in IRAC band 2.


\subsection{Outflows and the ambient medium}

The number of outflows identified here matches (to within 20\%) the
number of protostellar cores in L1448, L1455 and B1
\citep{hat07b}, though it is lower by a factor of 2 in NGC1333.  In this
densely-populated region this discrepancy is partly due to H$_2$
features being identified as part of a single flow, when in fact they
are associated with multiple flows.  Even so, the association of
protostellar cores with molecular (H$_2$/CO) outflows is clear.

It is also worth noting that the more massive clouds produce more
outflows in Perseus-West. In B1, NGC1333, L1448 and L1455 we identify
(at least) 8, 11, 4 and 5 outflows, respectively.  From near-IR
extinction mapping of Perseus, maps which cover an $A_{\rm v}$ range
of about 3-7, \citet{kir06} identify large scale structures which
they refer to as ``supercores''.  They stress that the bulk of the
cloud material is contained in these extensive, moderate-density
regions, rather than in the high-extinction cores not traced by the
extinction map. Kirk et al. measure masses of 441\Msolar , 973\Msolar
, 174\Msolar\ and 240\Msolar\ for the B1, NGC1333, L1448 and L1455
supercores, respectively.  Thus, in Perseus-West there is one molecular
outflow for every 44-88\Msolar\ of ambient material.  Roughly 50
generations of accretion/outflow activity would be needed to convert
all of this molecular material into solar-mass stars.

In some regions many of the jets appear to be roughly parallel, though
this common angle is not obviously related to the large-scale cloud
structure. L1448 is perhaps the best example, where the position
angles of the four outflows identified in Fig.~\ref{h2L1448} are
between 147\dg\ and 155\dg.  \citet{kir06} fit 2-dimensional Gaussians
to their supercores and, from their Fig.~5, we measure a supercore
position angle of 120\dg\ for L1448 .  In L1448, the outflows thus
appear to flow along the length of the supercore.  However, the two
parallel flows in the B1-ridge region (Fig.~\ref{h2B1r}; PAs of
124\dg\ and 130\dg ) are almost perpendicular to the elongated
B1-ridge supercore (PA $\sim$ 50\dg ).  The flows in B1, NGC1333 and
L1455 are more randomly orientated, although at least four of the
larger flows in NGC1333 are aligned roughly north-south, parallel with
the associated supercore (PA $\sim$ 10\dg ).  Even so, we conclude
that in general the outflows are randomly orientated and that
the large-scale could structure, as traced by the extinction mapping
of \citet{kir06}, has no obvious influence on flow orientations.


\subsection{Momentum injection and cloud support}

Molecular outflows from young stars inject turbulent energy and
momentum into the surrounding medium.  Over time these flows probably
provide turbulent support against cloud collapse and may even
hinder the star formation process. In low-mass star forming regions
usually only a dozen or so outflows are observed at any one time
(NGC1333 is a good example).  Here we investigate whether these
outflows provide sufficient kinetic energy to account for the observed
turbulence in the ambient medium.

As was noted by \citet{wal05a}, one can estimate the turbulent momentum
and energy in a star-forming region from the cloud mass, $M_{\rm c}$,
and velocity dispersion, $\delta V_{\rm tur}$, measured from
(sub)millimetre molecular lines:

\begin{itemize}

\item{For Perseus-West we again use cloud masses from \citet{kir06}.
The masses of their supercores (estimated from 2MASS extinction
mapping for $A_{\rm V}>$3) are listed in Sect.~4.4.  Although a
sizable fraction of the overall cloud mass is likely to be in diffuse,
lower-$A_{\rm V}$ regions, outflows are unlikely to impart much of
their momentum into these extended regions. Also, masses derived from
an extinction map are preferred over masses estimated from molecular
line observations, since the latter may suffer from depletion and/or
opacity effects.  And like submillimetre dust continuum maps,
optically thin molecular lines will only trace the denser core regions
($A_{\rm V}>$5)}.

\item{A turbulent velocity is difficult to generalise for each region.
The widths of molecular lines will result from turbulent and thermal
motions in the molecular gas, as well as macroscopic motions if many
independent cores are viewed along the same line of sight.  In nearby,
low-mass star forming regions the latter should not be a major
problem. However, due to opacity and depletion effects, different
molecular lines will trace different components of the ambient gas.
\citet{kir07} find that in Perseus N$_2$H$^+$ (1-0) lines are
broadened almost exclusively by thermal motions in dense cores;
C$^{18}$O (2-1) is more sensitive to lower density material in the
envelopes of dense cores (the C$^{18}$O transition has a critical
density of 10$^{3}$~cm$^{-3}$ and freezes out onto dust grains at
densities above 10$^{5}$~cm$^{-3}$). C$^{18}$O is therefore a better
tracer of the non-thermal motions in the interstellar -- and intercore
-- environment pervaded by molecular outflows. Across Perseus
C$^{18}$O profile widths vary by only a few tenths of a \kms .  From
Kirk et al. (2007) we therefore adopt a mean value of 0.6\kms\ for
$\delta V_{\rm tur}$ for all regions considered.  }

\end{itemize}

In Table~\ref{turb} we list the turbulent momentum $P_{\rm tur}\sim
M_{\rm c}\delta V_{\rm tur}$ and turbulent energy $E_{\rm
tur}\sim M_{\rm c}\delta V_{\rm tur}^2$ in each of the clouds observed
in Perseus.

The momentum and energy of an outflow from a young star is hard to
measure from infrared observations. Therefore, to estimate the
combined momentum and energy, $P_{\rm jets}$ and $E_{\rm jets}$, of
the outflows from the embedded YSOs in each supercore in Perseus, we
use the canonical values for the momentum flux for Class 0 and Class I
low mass protostellar outflows measured by \citet{bon96}, and multiply
these by the accretion lifetime of the embedded phase
\citep{wal05a,dav07}. The momentum and energy flux of an outflow is
thought to decrease with time, in line with decreasing accretion rates
\citep[e.g.][]{smi00}.  We therefore ignore the Class II/III outflow
phase.  If the momentum flux for the Class 0 and I phases is
 $\sim 6\times10^{-5}$~\Msolar \kms\ yr$^{-1}$ and 
 $\sim 4\times10^{-6}$~\Msolar \kms\ yr$^{-1}$ respectively, 
then for durations of $10^4$ and $10^5$ years, over its lifetime an
outflow from a low-mass protostar should inject $\sim$1.0~\Msolar
\kms\ of momentum into its surroundings.  The outflow kinetic energy
scales with the velocity; since most of the molecular material in
outflows is probably entrained in the relatively slow-moving wings of
bow shocks, then for a molecular flow velocity of $\sim$20\kms\
\citep[e.g.][]{kne00,hat07}, the kinetic energy supplied by a
typical flow would be $\sim4\times10^{37}$~J .  In Table~\ref{turb} we
thus multiply these canonical values by the number of observed H$_2$
flows in each region to get $P_{\rm jets}$ and $E_{\rm jets}$.

Although the combined energy of the outflows, $E_{\rm jets}$, matches
our estimates for the turbulent energy in each cloud, $E_{\rm tur}$, the
combined momentum, $P_{\rm jets}$, falls short of $P_{\rm tur}$ by at
least an order of magnitude in each region in Table~\ref{turb}. It has
been argued that a large fraction of the jet momentum may be carried
by a collimated, high-velocity jet
\citep{wal05a}.  However, the transport of this momentum component to
the surrounding cloud may be relatively inefficient, as the heavy jet
punches its way through the ambient medium.  The existence of
parsec-scale jets and the high proper motions of even the most distant
jet knots suggest that jets are relatively unhindered by their
surroundings.  Our assumption, that only the massive though
low-velocity molecular flow component contributes to the turbulent
motions in the ambient medium, is therefore probably still valid.

Instead, the turbulent momentum in the surrounding cloud may be built
up via many generations of outflows. The lifetime of a Giant Molecular
Cloud is 1-2 orders of magnitude longer than the duration of the Class
0/I phase, and Class 0, I and Class II YSOs clearly co-exist in many
star-forming regions \citep[e.g.][]{gut04,kum07,har07,gut08}.  Hence,
over the lifetime of a GMC, many generations of outflows may
contribute to the observed turbulent momentum in the
surrounding cloud.


\section{Conclusions}

We examine near- and mid-IR images of a $\sim$4 square degree
region in Perseus-West and compare these to published
catalogues of HH objects and moderately-extensive CO 3-2 maps.  We
find that in most cases molecular shock features are morphologically
the same at 2.12\mic\ and 4.5\mic , although the more embedded
portions of some flows are revealed only in the longer-wavelength
data. H$_2$ features are closely associated with high-velocity CO
lobes, in support of bow shock entrainment scenarios.  However, they
often do not coincide with their optical HH counterparts, because of
extinction effects or the discordant excitation requirements of each
tracer.

We use IRAC and MIPS photometry to identify and characterise the
outflow sources.  
In comparison to the full sample of YSOs in the c2d catalogue, the
molecular outflow sources possess extreme values of near-to-mid-IR
spectral index, $\alpha$, and highly reddened [3.6]-[4.5] and
[4.5]-[24] colours, as is befitting their extreme youth.

We find no correlation between H$_2$ flow length and
$\alpha$, nor between flow opening angle (as traced in H$_2$) and
$\alpha$. In general, the outflows are randomly orientated. However,
we do find a correlation between the number of outflows and the number
of protostellar cores in three of the four regions studies; in terms
of total cloud mass, there is one molecular outflow for every
44-88\Msolar\ of ambient material. The outflows may also be an
important source of turbulent energy in the interstellar medium.


\section*{Acknowledgements}

Warm thanks are due to Nanda Kumar and Tom Megeath for advice on
Spitzer data and YSO photometry, and to Jenny Hatchell and Lewis Knee
for making their CO maps of Perseus available to us.  We acknowledge
the Cambridge Astronomical Survey Unit (CASU) for processing the
near-IR data, and the WFCAM Science Archive in Edinburgh for making
the WFCAM data discussed here available to us.  The United Kingdom
Infrared Telescope is operated by the Joint Astronomy Centre on behalf
of the U.K. Particle Physics and Astronomy Research Council.  Some of
the WFCAM data reported here were obtained in Directors Discretionary
Time. This research made use of data products from the Spitzer Space
Telescope Archive. These data products are provided by the services of
the Infrared Science Archive operated by the Infrared Processing and
Analysis Center/California Institute of Technology, funded by the
National Aeronautics and Space Administration and the National Science
Foundation.  We have also made extensive use of the SIMBAD database,
operated at CDS, Strasbourg, France.



\begin{landscape}
\begin{table*}
\centering
\begin{minipage}{195mm}
\caption{Embedded YSOs and their molecular outflows in Perseus-West}
\begin{tabular}{@{}cccccccccccccc@{}}
  
\hline
Source$^a$ & H$_2$$^b$ & CO$^b$ & HH$^b$ & \multicolumn{6}{|c|}{Source Photometry and Spectral Index$^c$} & Length$^d$ & PA$^d$
& $\theta ^d$  & Region \\     
           & features    &        &        & [3.6] & [4.5] & [5.8] & [8.0] & [24] & $\alpha$  & (arcsec)   & (deg) & (deg)  \\ 
\hline
1 & 72,74 & y & 195 & 17.56 & 15.16 & 14.65 & 13.74 & 9.58 & 2.53 & 166 & 138 & 10 & L1448 \\
2 & 78-80,82-84 & y & 193,196,267 & - & 13.73 & 12.19 & 11.03 & 7.43 & 2.31 & 780 & 134 & - & L1448 \\
3 & 75-77 & y & 197 & 15.29 & 12.85 & 11.56 & 11.04 & - & 3.79 & 189 & 147 & 19 & L1448 \\
4 & 56-58 & y & 280,317,493,493 & 15.77 & 13.80 & 13.66 & 13.66 & 9.79 & 1.47 & 506 & 114 & 8 & L1451 \\
5 & 11-13,85? & n & n & 11.76 & 11.59 & 11.52 & 11.43 & 9.96 & -0.09 & 846 & 125 & 13 & L1455 \\
6 & 15-17,55 & y & 318 & 16.77 & 13.94 & 13.09 & 12.85 & 8.18 & 2.7 & 66 & 44 & 12 & L1455 \\
7 & 59 & y & n & 16.58 & 14.94 & 13.45 & 12.77 & 9.09 & 2.4 & 7 & 115 & - & L1455 \\
8 & 14,54 & ? & 422 & 9.89 & 9.49 & 9.10 & 8.68 & 8.27 & -0.19 & 37 & 179 & 8 & L1455 \\
9 & 66? & noOb & 741,744? & 15.28 & 16.18 & 15.28 & 14.92 & 11.92 & 0.68 & - & - & - & NGC1333 \\
10 & n & n & n & 14.61 & 13.57 & 12.86 & 12.56 & 11.84 & 0.16 & - & - & - & NGC1333 \\
11 & 20,24,46,71 & noOb & 15,338,339 & 13.24 & 10.87 & 10.59 & 9.31 & - & 3.16 & 461 & 140 & - & NGC1333 \\
12 & n & n & n & 18.54 & 17.68 & 17.02 & 16.47 & 12.81 & 1.71 & - & - & - & NGC1333 \\
13 & 61? & noOb & n & - & 18.10 & 17.08 & 16.42 & 14.42 & 0.91 & - & - & - & NGC1333 \\
14 & 22,23,62 & noOb & 340,343 & 17.11 & 15.15 & 15.25 & 15.30 & 10.47 & 1.63 & 99 & 98 & - & NGC1333 \\
15 & 21,31,32,41,42,45 & y & 13,341,342 & - & - & - & - & - & - & 405 & 12 & 11 & NGC1333 \\
16 & n & y & n & 12.80 & 11.15 & 10.44 & 9.96 & 7.32 & 1.4 & - & - & - & NGC1333 \\
17 & n & noOb & n & 11.67 & 11.57 & 11.77 & 11.75 & 11.29 & -0.74 & - & - & - & NGC1333 \\
18 & 26,27 & noOb & n & - & 19.03 & 18.30 & - & 13.29 & 2.18 & 127 & 98 & 10 & NGC1333 \\
19 & 38? & ? & 12? & - & - & - & - & - & - & - & - & - & NGC1333 \\
20a$^e$ & 35,60 & y & 7-11,753 & 8.90 & 11.67 & 7.53 & 19.77 & - & 0.37 & 114 & 132 & 17 & NGC1333 \\
20b$^e$ & 30,37,38 & y & 12 & - & - & - & - & - & - & 235 & 172 & 5 & NGC1333 \\
20c$^e$ & 19,28,34,43 & n & 14,344,350 & - & - & - & - & - & - & 987 & 10 & 4 & NGC1333 \\
21 & n & n & n & 17.13 & 16.15 & 16.01 & 15.96 & 11.65 & 1.55 & - & - & - & NGC1333 \\
22 & 29,33 & y & n & - & 18.92 & 18.62 & - & 12.57 & 2.63 & 89 & 34 & 9 & NGC1333 \\
23 & 39,40,44,48,65,67-70 & y & 5,347,761 & - & 14.12 & 13.80 & 13.89 & 9.09 & 1.9 & 412 & 159 & 12 & NGC1333 \\
24 & 36 & y & 6 & - & 14.80 & 14.92 & 15.33 & 9.50 & 2.18 & 27 & 48 & 58 & NGC1333 \\
25 & n & n & n & 17.41 & 13.58 & 15.31 & 15.51 & 12.44 & 0.72 & - & - & - & NGC1333 \\
26 & n & n & n & 19.32 & 16.69 & 16.52 & - & 12.67 & 1.8 & - & - & - & NGC1333 \\
28 & n & noOb & n & 16.34 & 15.15 & 14.34 & 13.83 & 11.73 & 1.06 & - & - & - & NGC1333 \\
29 & n & noOb & n & 12.22 & 12.23 & 11.98 & 11.42 & 10.59 & -0.32 & - & - & - & NGC1333 \\
32 & 49,50 & y & 773 & 18.19 & 16.73 & 16.77 & 16.59 & 13.38 & 1.12 & 291 & 124 & 17 & B1-ridge \\
33 & 2,3,4 & y & 782? & - & 15.37 & 15.21 & 15.92 & 13.50 & 0.09 & 217 & 130 & 12 & B1-ridge \\
35 & n & y & n & 18.28 & 16.32 & 15.52 & 14.80 & 11.16 & 2.17 & - & - & - & B1 \\
36 & n & n & n & - & - & - & - & 13.13 & - & - & - & - & B1 \\
37 & 10 & y & n & 14.46 & 12.03 & 11.46 & 11.21 & 8.31 & 1.9 & 13 &  90 & 53 & B1 \\
38 & 5-8 & y & n & 18.23 & 12.24 & 12.43 & 11.38 & 9.22 & 2.65 & 190 & 104 & - & B1 \\
39 & 9 & y & n & 14.75 & 13.22 & 12.78 & 12.59 & 10.76 & 0.67 & 45 & 62 & 18 & B1 \\
40 & 1 & ? & n & 16.16 & 14.13 & 14.05 & 13.62 & 8.31 & 2.54 & 3 & 109 & - & B1 \\
L1448-IRS1 & 73,81 & noOb & 194 & - & - & - & - & 8.04 & 0.38 & 153 & 115 & 18 & L1448 \\
LkH$\alpha$327 & 63 & noOb & 432,356? & 9.01 & 8.92 & 8.98 & 9.12 & 8.98 & -0.87 & 142 & 151 & 6 & B1 \\

\hline

\label{outflows}
\end{tabular}

\smallskip 
$^a$Embedded YSOs in Perseus-West; list compiled by \citet{jor07} from
Spitzer and SCUBA observations (see their Table 3). \\
$^b$H$_2$ features, molecular outflows, and HH objects associated with 
each YSO. See Table~\ref{jets} for additional details.\\
$^c$Source photometry obtained directly from the c2d catalogues.\\
$^d$Mean flow length, position angle and opening angle (average of two lobes).\\
$^e$Unresolved YSOs associated with three separate, clearly defined 
flows (see the online Appendix for details).\\

\end{minipage}
\end{table*}
\end{landscape}


\begin{table}
\centering
\begin{minipage}{82mm}
\caption{Turbulent momentum and energy in Perseus: clouds versus outflows}
\begin{tabular}{@{}lccccc@{}}
\hline
Region & $P_{\rm tur}$$^a$ & $E_{\rm tur}$$^a$  & N$^b$ & $P_{\rm jets}$$^c$ & $E_{\rm jets}$$^c$ \\
 or    & (\Msolar          & ($\times10^{38}$   &       & (\Msolar           & ($\times10^{38}$   \\
supercore &       \kms )   &                 J) &       &           \kms )   &                 J) \\
\hline

B1         &   264        & 3.2      &   8   &   8           & 3.2 \\
NGC1333    &   584        & 7.0      &  11   &  11           & 4.4 \\
L1448      &   104        & 1.3      &   4   &   4           & 1.6 \\
L1455      &   144        & 1.7      &   5   &   5           & 2.0 \\ 

\hline
\label{turb}
\end{tabular}
\smallskip 
$^a$Turbulent momentum and energy associated with the molecular cloud.\\
$^b$Number of H$_2$ flows in the region.\\
$^c$Total momentum and energy associated with the ``N" molecular outflows 
    in each region. 
\end{minipage}
\end{table}


\appendix

\section[]{Discussion of outflow activity in Perseus-West}

In this section we present colour-composite images of each region
constructed from WFCAM and IRAC observations, with 2.12\mic ,
3.6\mic\ and 4.5\mic\ data occupying the blue, green and red
channels, respectively (Figs~\ref{colorB1}--\ref{colorNGC1333}). 
These show how UKIRT/WFCAM and 
Spitzer/IRAC data complement each other in studies of outflows.  H$_2$
features appear pink/purple, since line-emission is in
almost all cases detected at 2.12\mic\ and 4.5\mic . 
Higher-excitation H$_2$ knots will appear slightly bluer, since
the WFCAM filter covers a rotation-vibrational line while the IRAC
band 2 covers only pure-rotational transitions (discussed further in
Sect.~4.3).  The outflow
sources, many of which are not visible in the near-IR, are in
most cases detected in the IRAC bands. The most embedded sources
appear red in the colour images, as do the most deeply embedded
outflow features that are not detected at 2.12\mic .  

These WFCAM/IRAC colour images are preferable to figures created from
three of the four IRAC bands since one avoids using the less sensitive
and relatively poor resolution IRAC bands 3 and 4 data: in colour
composites made solely from IRAC data (e.g. 3.6/4.5/8.0\mic) faint,
embedded sources may be undetected at the longer wavelength and thus
will appear green, similar to H$_2$ line-emission features that
are bright in IRAC band 2. The higher resolution WFCAM \htwo\ data
also help to ``sharpen'' the outflow features, producing an effect
similar to unsharp-masking.

Below we discuss the outflows in each region separately, relating
the H$_2$ features detected at 2.12\micron\ and/or 4.5\micron\ with
the high-velocity CO outflow lobes displayed in
Figs.~\ref{h2B1}--\ref{h2NGC1333}, the HH objects listed in the
literature, and the protostars identified from Spitzer photometry
\citep{eva05,jor07}.  Note that embedded YSOs from the list presented
by \citet{jor07} are referred to as YSO~1, YSO~2, etc., while sources
from the c2d catalogue are identified by their RA and Dec, and are
prefixed by ``J''.  Alternative source names are obtained from Table
A1 of \citet{hat05} and references therein; NGC1333-IRAS source names
are taken from
\citet{jen87}.


\begin{figure*}
\caption[] {{\bf SEE GIF} UKIRT \htwo\ (blue) - Spitzer 3.6\mic\ (green) - Spitzer
4.5\mic\ (red) colour composite image of B1.  H$_2$ features are numbered; 
embedded YSOs from \citet{jor07} are labelled.}
\label{colorB1}
\end{figure*}

\begin{figure*}
\caption[]
{{\bf SEE GIF} UKIRT \htwo\ (blue) - Spitzer 3.6\mic\ (green) - Spitzer 4.5\mic\
(red) colour composite image of B1 Ridge (west). H$_2$ features are numbered; 
embedded YSOs from \citet{jor07} are labelled.}
\label{colorB1r}
\end{figure*}

\begin{figure*}
\caption[]
{{\bf SEE GIF} UKIRT \htwo\ (blue) - Spitzer 3.6\mic\ (green) - Spitzer 4.5\mic\
(red) colour composite image of L1455. H$_2$ features are numbered; 
embedded YSOs from \citet{jor07} are labelled.}
\label{colorL1455}
\end{figure*}

\begin{figure*}
\caption[] {{\bf SEE GIF} UKIRT \htwo\ (blue) - Spitzer 3.6\mic\ (green) - Spitzer
4.5\mic\ (red) colour composite image of L1448. H$_2$ features are numbered; 
embedded YSOs from \citet{jor07} are labelled.}
\label{colorL1448}
\end{figure*}

\begin{figure*}
\caption[]
{{\bf SEE GIF} UKIRT \htwo\ (blue) - Spitzer 3.6\mic\ (green) - Spitzer 4.5\mic\ (red) 
colour composite image of NGC1333. H$_2$ features are numbered; 
embedded YSOs from \citet{jor07} are labelled.}
\label{colorNGC1333}
\end{figure*}


\subsection{B1}

We identify six outflows in B1. All six are driven by sources that are
identified by c2d and all except LkH$\alpha$327 are in the catalogue
of \citet{jor07}. (There is potentially a seventh outflow that may
have a source that is undetected by Spitzer between B1 and HH~429 -
see below.)

The deeply embedded Class 0 source YSO~38 (SMM~1, B1-c) is associated
with a well-defined bipolar CO outflow and the H$_2$ features 5 and
6-8 (Fig.~\ref{h2B1}).  The H$_2$ data show that the outflow extends
well beyond the bounds of the CO map. In our colour WFCAM/Spitzer
image (Fig.~\ref{colorB1}), an `S-shaped', precessing flow is evident;
the flow is traced back to the source in the longer wavelength
4.5\mic\ emission.  YSO~38 clearly drives knot 5 and probably much of
object 6-8 \citep[see also][]{wal05b}. Parts of the 6-8 complex could
be powered by J033309.3+311012 (the YSO situated midway between
features 6 and 7 in Fig.~\ref{h2B1}), though this source has a flat
spectral index ($\sim$0.08) and only a moderately red [3.6] - [4.5]
colour ($\sim$0.80; see e.g. Fig.~\ref{c2dcmag}). This source may
instead be a bright outflow feature misidentified by c2d as a YSO
candidate (discussed further in Sect.~4.3).

The overall morphology of the H$_2$ complex 6--8 also adds to the
confusion. Its east-facing bow shape seems to suggest movement toward
the east. 6--8 could conceivably be a bow shock in the counter-lobe of
a flow that drives the bright, opposite-facing HH object HH~429 (H$_2$
knot 53; see Walawender et al. 2005b - Fig.~30).  HH~429 is
17.5\arcmin\ due west of feature 53.

To the south of YSO~38, the nebulous, Class I source YSO~37
(Figs.~\ref{h2B1} and \ref{colorB1}) powers knot 10 to the west and a
red-shifted CO lobe to the east. This east-west orientation supports
the inclusion of HH~787, 785, 781 (and possibly HH~429) to the west
and HH~789 to the east in another parsec-scale flow
\citep{wal05b}.

The blue-shifted CO emission to the southwest of YSO~37
is probably associated with a bipolar flow from
YSO~35 (SMM3), although no H$_2$ emission is detected from this flow
at 2.12\mic .

YSO~39 (B1-b, SMM2,3) comprises two Class 0 sources
\citep[B1-bN and B1-bS;][]{hir99}. YSO~39 is associated with a
bright, knotty H$_2$ flow 9 to the southwest, and a blue-shifted CO
lobe to the northeast. 

YSO~40 is a nebulous infrared source tentatively associated with
blue-shifted CO emission and faint H$_2$ line emission (feature 50). The
source has a relatively steep spectral index, as expected for an
embedded outflow source ($\alpha$ = 2.54, [3.6]-[4.5] = 2.02).  It is
possibly associated with the bright HH features HH~431,433 and the jet
HH~790 \citep[][see their Fig.~32]{wal05b}.

Finally, LkH$\alpha$327, the brightest near-IR source in B1, drives a
northwest-southeast outflow and excites HH~432 and
H$_2$ feature 63. \citet{wal05b} suggest that HH~430 and HH~788 trace the
extended regions of the northern lobe, with HH~432 and 63 (their
MH~7) tracing the southern lobe.

\subsection{B1 Ridge}

The ridge of molecular gas running from B1 to L1455 contains two J\o
rgensen YSOs that are clearly driving molecular outflows
(Figs.~\ref{h2B1r}). YSO~33 (SMM 76, IRAS 03292+3039) is associated
with H$_2$ features 2, 3 and 4, while YSO~32 (SMM~77, IRAS~03282+3035)
powers H$_2$ features 49 (HH~773) and 50. Bow shock 49, first noted by
\citet{bal93}, is bright in \htwo\ (Fig.~\ref{colorB1r}) but only
marginally detected in the optical, as expected for embedded flows
from Class 0 objects.  The YSO~33 flow has no known HH counterparts.

51 and 52 are H$_2$ features located southwest along the B1 ridge (see
Table~\ref{jets} for details). The area has several Spitzer-identified
YSOs and red sources, along with several HH objects which include
HH~368-372, 427, 428, 769, and 792. However, none of the YSOs are in
the catalogue of \citet{jor07}
\citep[even though SCUBA data do exist for this region:][]{kir06}.
The HH objects and H$_2$ features do not have any clearly identifiable
sources, although H$_2$ features 51, 52 and IRAS~03272+3013 do lie on
the same axis.

\subsection{L1455 and L1451}

In the L1451/L1455 region in Fig.~\ref{h2L1455} we identify five H$_2$
outflows. At least four of these are driven by embedded YSOs
\citep{jor07}. 

The most prominent outflow emerging from the L1455 core includes
HH~279, H$_2$ knots 11-13 and possibly HH~423 and H$_2$ feature 85 in
a parsec-scale flow (see Table~\ref{outflows}). Class I objects YSO~5
(SMM~39) and YSO~8 (RNO~15) are candidate sources, although neither is
associated with high-velocity CO (Fig.~\ref{h2L1455}). The c2d source
J032719.0+301718 (coincident with HH~279) has a red [3.6]-[4.5] colour
of 1.55, though it also possess a negative spectral index (-0.46) and
therefore could again be a H$_2$ knot mis-identified as a YSO. YSO~8
may instead drive a north-south flow that includes H$_2$ knots 14, 54
and weak blue-shifted CO emission to the south.

Between YSOs 5 and 8 in the southeast corner of Fig.~\ref{h2L1455},
YSO~6 (SMM~35, RNO15-FIR, IRAS~03245+3002) powers H$_2$ features
15-17, 55 and a precessing bipolar CO outflow
\citep{dav97}.  YSO~7 (SMM~36) drives a
second, orthogonal CO flow that includes the faint H$_2$ feature
59. The blue lobe of this CO outflow may extend $\sim$1\arcmin\
northwest of knots 15 and 16, although no H$_2$ features are detected in
this extension. In the colour-composite image in Fig.~\ref{colorL1455}
YSOs 6 and 7 are obvious as the reddest sources in the region.

To the west of L1455, toward L1451, there is a parsec scale flow
driven by YSO~4 (SMM~80).  The flow is associated with H$_2$ features
56, 57 and 58 (HH~492, 493 and 317, respectively) to the west, and the
long chain of optical knots that comprise HH~280 to the east
\citep{wal04}. Red-shifted CO extends towards this eastern shock
feature.

\subsection{L1448}

There are four outflows identified in Fig.~\ref{h2L1448}. The reddened
source YSO~1 (SMM~30, IRAS~03222+3034, L1448-IRS2) drives a
spectacular bipolar flow associated with H$_2$ features 72 and 74
(Fig.~\ref{colorL1448}; see also Eisl\"offel 2000). Additional bright
line emission features are seen to the northwest of YSO~1, although
these are probably associated with a second bipolar flow from
L1448-IRS1 (RNO~13).  L1448-IRS1 is not catalogued
as a YSO by J\o rgensen. It has a spectral index of 0.38. It is
detected only in the Spitzer MIPS 1 band in the c2d catalogue and is
classified as ``star+dust''. The lack of a detection in the IRAC bands
is probably due to saturation as IRS1 is quite bright (see
Fig. \ref{colorL1448}).  H$_2$ knot 81 and HH~194 occupy the counter-lobe of
this flow. 

The parallel outflows from YSO~2 and YSO~3 (SMM~28/L1448N and SMM~29/L1448C,
respectively) have been studied extensively in the past
\citep[e.g][and references therein]{bac90,dav95,eis00}. The flow from YSO~2 includes
H$_2$ features 82 to the southeast and 78, 79, 80, 83 and 84 to the
northwest. These features increase the length of this northwestern
flow lobe to 25.5\arcmin\ (2.2~pc). The YSO~3 flow is associated with
H$_2$ object 77 in the south and the spectacular, curving H$_2$
flow 76 to the north.  Feature 75 is an unusual group of
knots to the northwest of 76: we tentatively associate these with the
curving 76 outflow, though on morphological grounds one might regard
75 and 78 as bow shocks and counter-bow shocks in a bipolar flow
centred midway between these two objects.  However, no candidate
source is found in this region.  

Finally, we bring the reader's attention to a highly reddened source
32\arcsec\ south of YSO~3.  This nebulous source can be seen clearly
in the colour image in Fig. \ref{colorL1448}.  The source,
J032539.6+304336, is classified by c2d as ``cup-down'', i.e. it has
been detected in only 3 IRAC bands with a magnitude at 4.5\mic\ that
is brighter than at 3.6\mic\ and 5.8\mic\ (possibly due to unresolved
line emission). However, close examination of the Spitzer IRAC images
reveals that the source is in fact detected in all four IRAC bands.
From our own photometry, measured from IRAC BCD images using a
2\arcsec\ aperture, we measure a [3.6]-[4.5] colour of 2.2 (typical
for outflow sources) and a spectral index, $\alpha$, of 2.33 (again
typical for embedded Class 0/I outflow sources).  Hence, this source
could be contributing to the outflow activity in this region.

\subsection{NGC1333}

The outflows in NGC1333 have been studied in considerable detail,
though the WFCAM images presented here are the most extensive H$_2$
images to date and as a result reveal new features.  In
Fig.~\ref{h2NGC1333} and \ref{h2NGC1333S} we mark 11 outflows. Some of
these have previously been noted by \citet{hod95}.  All 11 are
associated with
\citet{jor07} YSO sources. 

\subsubsection{The SVS-13 SMM core}

There are three major outflows emanating from the general area of
SVS~13 and YSO~20. YSO~20 is identified by
\citet{jor07} as the centre of the SCUBA submillimeter core
rather than a YSO. This is because YSO~20 was identified using only
SCUBA data and so no c2d YSO position is available. This
submillimeter core  contains
5 radio continuum sources which \citet{rod99} identify as
YSOs.

Rather than speculate on the source of each of the flows emerging from
the SVS~13 core, in Table~\ref{outflows}, we identify the sources
according to their alignment with outflow features and/or CO flow
lobes: YSO~20a is deemed to be the driver of the well-know bipolar CO
outflow containing HH~7-11; YSO~20b is the driver of the north-south
flow at PA $\sim$ 170\dg\ which includes H$_2$ features 30 in the
south, 37 in the north and possible feature 38 (HH~12); YSO~20c is
then the driver of the lone south lobe at PA $\sim$ 10\dg . This
latter chain of H$_2$ features extends at least 16.5\arcmin\ (1.4~pc)
to the south and includes H$_2$/HH objects 34, 28, 43, HH~14 and
19. There is no obvious counter-lobe in our H$_2$ image, although
HH~754/755 do lie north along an axis drawn through the southern lobe
\citep{wal05b}.  There is also no clear evidence of an outflow in the
CO data of \citet{kne00}.
Our colour-composite image in Fig. \ref{colorNGC1333} reveals a
bright, nebulous source roughly 1\arcmin\ south of SVS~13 that could
be associated with the YSO~20c flow.  The source is associated with a
conical reflection nebula that opens out toward the south.

\subsubsection{IRAS~2A and 2B}

YSOs 15 and 16 are better known as IRAS~2A (SMM~44) and
IRAS~2B. YSO~15 is associated with a SCUBA core and VLA radio source
\citep{rod99}.  In the colour image in Fig. \ref{colorNGC1333} YSO~15
appears highly reddened and elongated along the identified flow
direction.  It is clearly associated with H$_2$ knots 31 and 32,
and the bipolar CO flow evident in Fig.~\ref{h2NGC1333}. However, the
YSO~15 ``core'' may also be associated with a second, orthogonal flow,
which we label 45.  This bipolar flow can also be seen clearly in
the CO data. The north-south flow from YSO~15 is a large outflow that
influences HH~12 in the north and extends at least as far down as
HH~13 and H$_2$ features 41/42 in the south. Its axis also lines up with
knot 18 (HH~746) $\sim$25\arcmin\ to the south of YSO~15. The extent of
this flow lobe could therefore be at least 2.2~pc.

There are no H$_2$ features that are obviously associated with YSO~16
(IRAS~2B).

\subsubsection{HH~12}

HH~12 is a complex cluster of knots about 5\arcmin\ north of SVS~13
and HH~7-11. The bright infrared source (IRAS~6) associated
with HH~12 sits atop a column of dense, molecular gas; the source is
flagged as a YSO by \citet{jor07} based on its association with a
SCUBA core (SMM~45).

We label the H$_2$ features around HH~12 as 37/38. Feature 37 is a faint
trail of emission leading from near SVS~13 to the head of HH~12; 38
is the complex group of bright knots that make up the head of
HH~12. Part of HH~12 could be associated with the H$_2$ flow 31/32
from YSO~15/IRAS~2A; HH~12 is clearly aligned with the
well-defined axis of this bipolar outflow.  However, the spray of
knots to the west of YSO~19, coupled with the overlapping blue-shifted
and red-shifted CO observed towards this source,
point to an outflow inclined towards the line of sight that is driven
by a source within HH~12. \citet{rod99} find 2 radio continuum sources
that coincide with the YSO~19 core and there are three flat-spectrum
Spitzer YSOs about 1\arcmin\ to the southwest of YSO~19 that could be
contributing to the outflow activity, although one or more of them
could again be an H$_2$ knot misidentified as a c2d YSO.

\subsubsection{Other outflows in NGC1333}

Northeast of SVS~13, H$_2$ feature 36 (HH~6) and its associated CO
outflow are driven by either YSO~23 or YSO~24 (IRAS~7, SMM~46). There
is also an extensive flow perpendicular to 36 that is likewise driven
by either YSO~23 or 24. This large flow runs from H$_2$ knot 44/HH~5
in the south to feature 68 in the north (see
Fig.~\ref{colorNGC1333}). Feature 25/HH~18 and HH~761 may also be part
of this flow.  In Tables \ref{outflows} and \ref{jets} YSO~23 is
identified as the source of the extensive northeast-southwest jet and
YSO~24 as the source of HH~6, because YSO~24 is slightly better
aligned with the central axis of the feature 36/HH~6 cluster. However,
the reverse cannot be ruled out.

NGC1333 also contains two ``S-shaped'' outflows, driven by YSOs 11 and
14 (Figs.\ref{h2NGC1333} and \ref{h2NGC1333S}). YSO~11 (SMM~49) drives
H$_2$ features 20, 46 and possibly 71/HH~338 (8.1\arcmin\ to the
northwest), and feature 21 and possibly even 24/HH~15 (7.3\arcmin\ to
the southeast). YSO~14 drives H$_2$ knots 22/23/62 and HH~340/343 in
another S-shaped flow \citep[for a detailed discussion of this outflow
see][]{hod05}. Both flows are remarkably similar to the bending jet
from YSO~38 in B1 (Fig.~\ref{colorB1}).

3\arcmin\ southeast of SVS~13, YSO~22 (SMM~41) drives a well-defined
bipolar CO outflow that includes H$_2$ features 33 and 29
(Fig.~\ref{h2NGC1333}). To the southwest, this flow may be interacting
with the flow from the SVS~13 core and to the northeast with both the
HH~7-11 outflow and the southern lobe of the YSO~23 jet.

Finally, near the bottom of Fig.~\ref{h2NGC1333}, YSO~18 (SMM~65)     
drives an east-west flow associated with H$_2$ bows 26 to the east and
28 to the west. Unfortunately the CO map of \citet{kne00} does not extend
sufficiently southward to contain YSO~18 and its outflow, or indeed
any of the other southern flows labelled in Fig.~\ref{h2NGC1333S}.

 
\clearpage
\begin{table*}
\centering
\begin{minipage}{175mm}
\caption{H$_2$ features in Perseus-West}
\begin{tabular}{@{}cccccccllc@{}}
  
\hline
H$_2$  &  RA$^a$    &  Dec $^a$    & Source$^b$     & Length$^c$ & PA$^d$    & HH$^e$ & Comment        & Other$^f$        & Figures$^g$  \\     
feature & (J2000.0) & (J2000.0)    & candidate      & (arcsec)   & (deg)     &        &                & identifiers      &        \\ 
\hline

1  & 3:33:27.417 & 31:07:10.06  & 40             & 3          & 109       &        &                & MH 12            & \ref{h2B1},\ref{colorB1}\\
2  & 3:32:11.860 & 30:50:51.55  & 33             & 243        & -49       &        &                &                  & \ref{h2B1r}\\
3  & 3:32:26.889 & 30:48:14.22  & 33             & 191        & 127       &        &                &                  & \ref{h2B1r}\\
4  & 3:32:38.992 & 30:44:16.92  & 33             & 425        & 140       &        &                &                  & \ref{h2B1r}\\
5  & 3:33:33.539 & 31:08:40.17  & 38             & 214        & 104       &        &                & MH 1             & \ref{h2B1},\ref{colorB1}\\ 
6  & 3:33:11.802 & 31:09:58.03  & 38             & 92         & -71       &        &                & part of MH 2     & \ref{h2B1},\ref{colorB1}\\
7  & 3:33:07.400 & 31:10:16.64  & 38             & 154        & -71       &        &                & part of MH 2     & \ref{h2B1},\ref{colorB1}\\
8  & 3:33:07.575 & 31:09:49.25  & 38             & 165        & -76       &        &                & part of MH 2     & \ref{h2B1},\ref{colorB1}\\
9  & 3:33:18.648 & 31:07:14.35  & 39             & 45         & -118      &        &                & MH 5             & \ref{h2B1},\ref{colorB1}\\
10 & 3:33:15.769 & 31:07:54.69  & 37             & 13         & -90       &        &                & MH 4             & \ref{h2B1},\ref{colorB1}\\

11 & 3:27:01.380 & 30:18:27.11  & 5              & 606        & -60       &        & same flow as 12? &                & \ref{h2L1455},\ref{colorL1455}\\
12 & 3:27:16.405 & 30:17:14.44  & 5              & 408        & -55       & 279    &                &                  & \ref{h2L1455},\ref{colorL1455}\\
13 & 3:27:20.454 & 30:17:15.97  & 5              & 258        & -47       &        &                &                  & \ref{h2L1455},\ref{colorL1455}\\
14 & 3:27:47.874 & 30:12:24.62  & 8              & 44         & -12       &        &                &                  & \ref{h2L1455},\ref{colorL1455}\\
15 & 3:27:38.776 & 30:12:56.34  & 6              & 25         & -136      &        &                &                  & \ref{h2L1455},\ref{colorL1455}\\
16 & 3:27:37.773 & 30:12:46.72  & 6              & 25         & -136      &        & same as 6      &                  & \ref{h2L1455},\ref{colorL1455}\\
17 & 3:27:42.044 & 30:13:40.50  & 6              & 63         & 33        &        &                &                  & \ref{h2L1455},\ref{colorL1455}\\

18 & 3:28:32.601 & 30:50:30.12  & -              & 57         & 161       & 746?   & 98\arcsec\ from HH   &            & \ref{h2NGC1333S}\\
19 & 3:28:49.421 & 31:00:11.13  & 20c            & 987        & -171      & 14?    & 32\arcsec\ from HH   &            & \ref{h2NGC1333S}\\
20 & 3:28:36.965 & 31:13:31.42  & 11             & 100        & -50       & 339?   &                &                  & \ref{h2NGC1333},\ref{colorNGC1333}\\
21 & 3:28:50.455 & 31:10:09.99  & 15             & 286        & -166      & 341?   &                &                  & \ref{h2NGC1333S}\\
22 & 3:28:39.820 & 31:05:46.92  & 14             & 79         & -83       & 340B   &                &                  & \ref{h2NGC1333S}\\
23 & 3:28:53.872 & 31:05:25.10  & 14             & 119        & 99        & 343    &                &                  & \ref{h2NGC1333S}\\
24 & 3:28:59.008 & 31:08:00.70  & 11             & 435        & 139       & 15     &                &                  & \ref{h2NGC1333S}\\
25 & 3:29:25.710 & 31:07:26.32  & -              & 26         & 134       & 18     &                &                  & \ref{h2NGC1333S}\\

26 & 3:29:10.978 & 31:11:44.34  & 18             & 135        & 97        &        &                & HL 8             & \ref{h2NGC1333},\ref{colorNGC1333}\\ 
27 & 3:28:51.779 & 31:12:23.14  & 18             & 118        & -80       &        &                & ASR 97           & \ref{h2NGC1333},\ref{colorNGC1333}\\
28 & 3:28:58.846 & 31:12:19.84  & 20c            & 223        & -166      & 344?   & 66\arcsec\ from HH   & ASR 98     & \ref{h2NGC1333},\ref{colorNGC1333}\\
29 & 3:29:06.977 & 31:12:15.71  & 22             & 88         & -147      &        &                & HL 3,5           & \ref{h2NGC1333},\ref{colorNGC1333}\\
30 & 3:29:05.543 & 31:12:51.03  & 20b            & 193        & 174       &        &                & HL 4             & \ref{h2NGC1333},\ref{colorNGC1333}\\
31 & 3:28:53.911 & 31:13:24.08  & 15             & 76         & -164      &        &                & ASR 49,71        & \ref{h2NGC1333},\ref{colorNGC1333}\\
32 & 3:28:56.054 & 31:15:47.67  & 15             & 74         & 10        &        &                & ASR 15,16        & \ref{h2NGC1333},\ref{colorNGC1333}\\
33 & 3:29:14.449 & 31:14:44.03  & 22             & 90         & 35        & 347?   &                & ASR 57           & \ref{h2NGC1333},\ref{colorNGC1333}\\
34 & 3:29:02.431 & 31:15:01.86  & 20c            & 54         & -169      &        &                & ASR 20,21        & \ref{h2NGC1333},\ref{colorNGC1333}\\
35 & 3:29:00.695 & 31:16:58.24  & 20a	         & 112        & -35       & 753    &                &                  & \ref{h2NGC1333},\ref{colorNGC1333}\\
36 & 3:29:12.865 & 31:18:44.88  & 24             & 27         & 48        & 6      &                &                  & \ref{h2NGC1333},\ref{colorNGC1333}\\ 
37 & 3:28:59.208 & 31:18:51.44  & 20b            & 182        & -10       &        &                &                  & \ref{h2NGC1333},\ref{colorNGC1333}\\
38 & 3:28:57.779 & 31:20:19.78  & 20b            & 265        & -10       & 12     &                &                  & \ref{h2NGC1333},\ref{colorNGC1333}\\
39 & 3:29:03.812 & 31:21:57.75  & 23             & 237        & -22       &        &                &                  & \ref{h2NGC1333},\ref{colorNGC1333}\\
40 & 3:29:00.448 & 31:24:04.77  & 23             & 369        & -21       &        &                &                  & \ref{h2NGC1333},\ref{colorNGC1333}\\

41 & 3:28:46.819 & 31:07:08.72  & 15             & 464        & -166      & 13     &                &                  & \ref{h2NGC1333S} \\
42 & 3:28:44.259 & 31:07:34.42  & 15             & 449        & -161      & 13?    & 50\arcsec\ from HH   &            & \ref{h2NGC1333S}\\
43 & 3:28:56.631 & 31:07:37.47  & 20c            & 504        & -170      &        &                &                  & \ref{h2NGC1333S} \\

44 & 3:29:20.190 & 31:12:49.95  & 23             & 352        & 160       & 5      &                &                  & \ref{h2NGC1333},\ref{colorNGC1333} \\
45 & 3:28:59.548 & 31:14:22.58  & 15             & 53         & 106       &        &                & HL 2             & \ref{h2NGC1333},\ref{colorNGC1333} \\
46 & 3:28:42.589 & 31:12:35.24  & 11             & 90         & 128       &        &                &                  & \ref{h2NGC1333},\ref{colorNGC1333}\\
47 & 3:29:27.277 & 31:13:30.71  & -              & 6          & 63        &        &                &                  & \ref{h2NGC1333}\\
48 & 3:29:10.873 & 31:19:53.97  & 23             & 85         & 0         &        &                &                  & \ref{h2NGC1333},\ref{colorNGC1333}\\

49 & 3:31:31.013 & 30:44:11.37  & 32             & 216        & 126       & 773    & HH faint in optical &             & \ref{h2B1r},\ref{colorB1r}\\
50  & 3:31:02.864 & 30:47:47.50  & 32             & 365        & -59       &        &                &                 & \ref{h2B1r},\ref{colorB1r} \\
51 & 3:30:33.338 & 30:27:51.29  & -              & 21         & 155       & 370    &                &                  & -- \\
52 & 3:30:45.780 & 30:29:19.27  & -              & 10         & 154       & 372    &                &                  & -- \\
53 & 3:31:47.973 & 31:10:00.60  & -              & 47         & 94        & 429    &                &                  & -- \\

54 & 3:27:47.978 & 30:11:44.37  & 8              & 30         & 180       & 422    &                &                  &  \ref{h2L1455},\ref{colorL1455}\\
55 & 3:27:44.810 & 30:14:15.03  & 6              & 107        & 44        & 318    &                &                  &  \ref{h2L1455},\ref{colorL1455}\\
56 & 3:26:49.129 & 30:14:56.63  & 4              & 165        & 105       & 493    &                &                  &  \ref{h2L1455},\ref{colorL1455}\\
57 & 3:26:56.873 & 30:14:14.70  & 4              & 263        & 106       & 317D   &                &                  &  \ref{h2L1455},\ref{colorL1455}\\
58 & 3:27:12.974 & 30:11:59.29  & 4              & 506        & 114       & 317    &                &                  &  \ref{h2L1455},\ref{colorL1455}\\
59 & 3:27:43.006 & 30:12:29.64  & 7              & 7          & 115       &        &                &                  &  \ref{h2L1455},\ref{colorL1455}\\

 \hline

\label{jets}
\end{tabular}
\end{minipage}
\end{table*}

\begin{table*}
\centering
\begin{minipage}{175mm}
\contcaption{}
\begin{tabular}{@{}cccccccllc@{}}
  
\hline
MHF  &  RA$^a$    &  Dec $^a$    & Source$^b$     & Length$^c$ & PA$^d$    & HH$^e$ & Comment          & Other$^f$      & Figure$^g$  \\ 
    & (J2000.0)  & (J2000.0)    & candidate      & (arcsec)   & (deg)     &        &                  & identifiers    &         \\ 
\hline

60 & 3:29:08.163 & 31:15:29.47  & SVS 13         & 116        & 119       & 7-11   &                & HL 7             &  \ref{h2NGC1333},\ref{colorNGC1333}\\
61 & 3:28:42.985 & 31:17:45.12  & 13             &            &           &        &                  &                &  \ref{h2NGC1333},\ref{colorNGC1333}\\
62 & 3:28:42.879 & 31:05:36.47  & 14             & 32         & 99        & 340A   & same flow as 22  &                &  \ref{h2NGC1333S}\\

63 & 3:33:35.820 & 31:08:45.76  & LkH$\alpha$327 & 142        & 151       &        &                  & MH 7           &  \ref{h2B1}\\
64 & 3:33:13.323 & 31:08:18.11  & -              & 5          & 90        &        &                  & MH 3           &  \ref{h2B1}\\

65 & 3:29:15.352 & 31:17:03.67  & 23             & 101        & 158       &        &                  &                &  \ref{h2NGC1333},\ref{colorNGC1333}\\
66 & 3:28:33.274 & 31:12:14.42  & -              & 8          & 163       & 744B   &                  &                &  \ref{h2NGC1333},\ref{colorNGC1333}\\
67 & 3:29:17.212 & 31:15:26.51  & 23             & 192        & 154       & 374?   &                  & HL 10,11       &  \ref{h2NGC1333},\ref{colorNGC1333}\\
68 & 3:28:57.640 & 31:25:32.45  & 23             & 472        & -22       &        &                  &                &  -- \\
69 & 3:29:03.949 & 31:23:02.04  & 23             & 295        & -17       &        &                  &                &  \ref{h2NGC1333},\ref{colorNGC1333}\\
70 & 3:29:07.627 & 31:21:04.69  & 23             & 170        & -14       &        &                  &                &  \ref{h2NGC1333}\\
71 & 3:28:12.876 & 31:19:43.71  & 11             & 486        & -40       & 338    &                  &                &  -- \\  

72 & 3:25:26.718 & 30:44:04.20  & 1              & 122        & 140       &        &                  &                &  \ref{h2L1448},\ref{colorL1448}\\
73 & 3:25:17.600 & 30:45:49.10  & L1448-IRS1     & 138        & 112       &        &                  &                &  \ref{h2L1448},\ref{colorL1448}\\
74 & 3:25:14.925 & 30:46:54.59  & 1              & 209        & -44       & 195    &                  &                &  \ref{h2L1448},\ref{colorL1448}\\
75 & 3:25:27.120 & 30:45:50.57  & 3              & 232        & -53       &        &                  &                &  \ref{h2L1448},\ref{colorL1448}\\
76 & 3:25:38.140 & 30:44:36.80  & 3              & 94         & -30       & 197    &                  &                &  \ref{h2L1448},\ref{colorL1448}\\
77 & 3:25:41.056 & 30:42:00.64  & 3              & 145        & 167       &        &                  &                &  \ref{h2L1448},\ref{colorL1448}\\
78 & 3:25:12.777 & 30:49:22.86  & 2              & 390        & -52       & 196    &                  &                &  \ref{h2L1448},\ref{colorL1448}\\
79 & 3:24:52.682 & 30:54:47.39  & 2              & 863        & -47       & 193    &                  &                &  \ref{h2L1448}\\
80 & 3:25:25.696 & 30:46:55.18  & 2              & 168        & -56       &        &                  &                &  \ref{h2L1448},\ref{colorL1448}\\
81 & 3:24:58.056 & 30:47:42.66  & L1448-IRS1     & 168        & -61       & 194    &                  &                &  -- \\
82 & 3:25:36.831 & 30:45:14.86  & 2              & 34         & 140       &        &                  &                &  \ref{h2L1448},\ref{colorL1448}\\
83 & 3:24:26.679 & 30:56:44.31  & 2              & 1128       & -53       &        &                  &                &  -- \\
84 & 3:24:02.837 & 31:00:25.00  & 2              & 1530       & -52       & 267    &                  &                &  -- \\

85 & 3:28:40.874 & 30:01:59.92  & 5              & 1087       & 130       &        &                  &                &  --\\

\hline

\end{tabular}
\smallskip \\
$^a$Position of the brightest knot within the H$_2$ feature or flow. \\
$^b$Most likely source. Numbers refer to Table 3 in \citet{jor07}. A dash means there is no obvious source candidate.\\
$^c$The distance from the source to the most distant knot. If no known source, entire length of all knots.\\
$^d$Position angle, measured east of north, of an axis drawn from the source to the feature.\\
$^e$Associated HH object (if any is known). \\
$^f$Previous names of H$_2$ features. ASR refers to sources from \citet{asp94}; HL refers to sources from \citet{hod95}; 
MH refers to sources from \citet{wal05b} \\
$^g$Figures in which the feature appears. \\
\end{minipage}
\end{table*}

\bsp
\label{lastpage}
\end{document}